\documentclass[traditabstract]{aa} 
\usepackage{graphicx}
\usepackage{txfonts}
\usepackage{natbib}
\usepackage{longtable}
\bibpunct{(}{)}{;}{a}{}{,} 
\begin{document}
   \title{A comprehensive view of the Virgo Stellar Stream}


   \author{Sonia Duffau \inst{1,2}
      \and A. Katherina Vivas \inst{3}
	  \and Robert Zinn \inst{4}
	  \and Ren\'e A. M\'endez \inst{5}
	  \and Mar{\'\i}a Teresa Ruiz \inst{5}
          }
   \institute{Zentrum f\"{u}r Astronomie der Universit\"{a}t Heidelberg, Landessternwarte, K\"{o}ningstuhl 12, D-69117, Heidelberg, Germany\\
              \email{sduffau@lsw.uni-heidelberg.de}
   \and
       Zentrum f\"{u}r Astronomie der Universit\"{a}t Heidelberg, Astronomisches Rechen-Institut, M\"{o}nchhofstra\ss{}e 12-14, D-69120 Heidelberg, Germany\\
              \email{sduffau@ari.uni-heidelberg.de}
     \and
            Cerro Tololo Interamerican Observatory, Casilla 603, La
            Serena, Chile\\
            \email{kvivas@ctio.noao.edu}
     \and   
	      Department of Astronomy, Yale University, P.O. Box 208101, New Haven, CT 06520-8101, USA\\
	      \email{robert.zinn@yale.edu}
	  \and
	     Departmento de Astronom{\'\i}a, Universidad de Chile, Casilla 36-D, Santiago, Chile\\
	     \email{rmendez@u.uchile.cl, mtruiz@das.uchile.cl}
	     }

\abstract {To explore the complex halo substructure that has
been reported in the direction of the Virgo constellation, radial
velocities and metallicities have been measured for 82 RR Lyrae stars
(RRLS) that were identified by the QUEST survey.  These stars are
distributed over 90 sq. deg. of the sky, and lie from 4 to 23 kpc
from the Sun.  Using an algorithm for
finding groups in phase space and modeling the smooth halo component in the region, we identified the 5 
most significant RRLS groups, some of which were previously known or
suspected. We have examined the SEKBO and the Catalina catalog of RRLS (with available spectroscopic 
measurements by Prior et al. 2009, and Drake et al. 2013), as well as
the bright QUEST RRLS sample (Vivas et al. in prep.), 
the catalogs of blue horizontal
branch (BHB) stars compiled by Sirko et al (2004) and Brown et al (2008, 2010) and the catalog of 
Red Giant stars from the Spaghetti survey,
for stars that may be related to the QUEST RRLS groups. The
most significant group of RRLS is the Virgo Stellar Stream (VSS, first
reported by Duffau et al 2006) identified here as group A,
which is composed of at least 10 RRLS and 3 BHB stars.  It has a mean distance of 19.6 kpc and a mean radial
velocity $V_{\rm gsr} = 128$ ${\rm km~s}^{-1}$, as estimated from its RRLS members.  With the revised
velocities reported here, there is no longer an offset in velocity
between the RRLS in the VSS and the prominent peak in the velocities
of main-sequence turnoff stars reported by Newberg et al (2007) in
the same direction and at a similar distance (S297+63-20.5). 
The location in phase space of two other groups (F and H)
suggests a possible connection with the VSS, which cannot be discarded
at this point, although the turnoff colors of the VSS and group H, as
identified from Newberg et al. 2007, suggest they might be composed of
different populations. Two more groups, B and D, are found at mean distances of 19.0
and 5.7 kpc, and mean radial velocities of $V_{\rm gsr} = -94$ and $32$ ${\rm km~s}^{-1}$.
The latter is the more numerous in terms of total members, as well as
the more extended in RA.   
A comparison with the latest
model of the disruption of the Sagittarius dwarf, indicates that none
of the above groups is related to it.  
Rather than being the result of a single accretion event, the excess
of stars observed in Virgo appears to be composed of several halo substructures along the same line of sight.  }

\keywords{Stars: variables: RR Lyrae -- Galaxy: abundances -- Galaxy: halo --
	    Galaxy: kinematics and dynamics -- Galaxy: structure }

\maketitle

\section{INTRODUCTION \label{sec-intro}}

Over the last decade, a number of simulations of galaxy formation that
are based on the popular $\Lambda$CDM hierarchical picture 
have predicted that the halos of disk galaxies should contain numerous
substructures that are the debris from 
accreted dwarf galaxies \citep{bullock05,cooper10}.  Over the same
period, the evidence 
that the Milky Way has indeed accreted dwarf galaxies has grown
considerably \citep[e.g.][]{bell08,helmi11,xue11}. 
 While there is no doubt that accretion has occurred, it is not yet
 firm that the number and 
the properties of the observed substructures are consistent with those
simulations.  
To answer this question, it is not only necessary to find the
substructures but also 
to characterize them in sufficient detail that the masses of the dwarf
galaxies 
and the times of their accretion can be estimated.  This paper reports 
a more thorough description of halo substructure in the direction of the constellation Virgo.

After the Sagittarius (Sgr) tidal streams \citep[e.g.][]{majewski03},
the Virgo region contains the most obvious overdensity of stars in
the sky explored so far by large scale surveys. Discovered as an
overdensity of RR Lyrae stars (RRLS) in the QUEST survey
\citep{vivas01,vivas02,vivas06,duffau06} and of main-sequence turn-off
stars \citep{newberg02} in the SDSS, it is one of the most noticeable features in
the Field of Streams \citep{belokurov06}. Based in main sequence stars
from SDSS, \citet{juric08} estimated that the substructure, covers $\sim 1000$
sq deg of the sky. More recently, \citet{bonaca12} suggested the
feature may span up to $\sim 3000$ sq deg. \citet{juric08} suggested that
the Virgo overdensity (VOD) was produced by the merger of a low surface brightness
galaxy with the Milky Way (see also \citet{carlin12}). Spectroscopic observations have
revealed, however, a complex system of substructures in the kinematic
distribution of stars
\citep{duffau06,newberg07,vivas08,starkenburg09,prior09a,brink10,casey12}. Furthermore,
it appears that the strengths of these features depend on the tracer used.  For example, the strongest
peak in the velocity distribution of RRLS and turnoff stars
in the region is found near $V_{\rm gsr}=120$ ${\rm km~s}^{-1}$
\citep{duffau06,newberg07,prior09a}, while main sequence and K giants
show a peak at $V_{\rm gsr} = -80$ ${\rm km~s}^{-1}$
\citep{brink10,casey12}.

In their study of substructures in high resolution simulations of
galactic halos from the Aquarius project, \citet{helmi11} found that
debris from massive progenitors often look like diffuse substructures
on the sky.  They point out that because of the large-scale
structure present in the Universe when galaxies began to form, the accretion of galaxies was not random in direction, but was instead along preferred directions.
This infall pattern causes the streams and substructures of different progenitors to overlap.  This scenario, rather than a single massive progenitor, may
be the explanation for the substructures in Virgo. To complicate the
picture, the region of the VOD is not far from the leading tail of the
Sgr dSph galaxy.  Although the bulk of the Sgr stars lie much farther
away, at 50 kpc, it has been suggested that stars which became unbound
in previous passages of the galaxy may lie at much closer distances
\citep{martinez07,prior09a,law10}.

Although the distance to the Virgo overdensity is relatively small, the spatial
distribution of stars alone does not provide a clear picture of the complex substructure.
The densest part of the RRLS overdensity lies at 19 kpc
\citep{vivas06}, but \citet{juric08} suggested that the VOD covered a
range of distances from $\sim 5$ to 20 kpc. Velocities are key to
separate the different accretion events and establish the relationship
between them, if any. 

In this investigation, we obtained spectroscopy of RRLS present in the
QUEST survey in the Virgo region between 4 and 23 kpc, 
with the goal of filling the gap between the
earlier investigations by \citet{duffau06} and \citet{vivas08} who
observed QUEST RRLS in ranges 18-20 kpc and $< 12$ kpc, respectively.
The sample of RRLS presented here, which is a combination of new
observations and an updating of previous ones, includes 87\% of the
QUEST RRLS in an area of almost 90 sq degrees of the sky. The
relative errors of the RRLS distances are about 7\% (Vivas and Zinn
2006), which is superior to most other halo tracers (e.g., $\sim 15\%$
for red giants, Starkenburg 2009) and comparable to that obtained for
BHB stars, their nearest rival.  The pulsations of RRLS make the
determination of their systemic radial velocities more difficult than
for other stars, but even using small numbers of observations of
modest precision, it is possible to obtain precisions of $\sim15-20$ 
${\rm km~s}^{-1}$ (e.g. Layden 1994, Vivas et al. 2005, Prior et al. 2009).
Since the line-of-sight (los) velocity dispersion of the halo is $\sim
115$ ${\rm km~s}^{-1}$ over the distance range considered here \citet{brown10}, this
precision is adequate for identifying the velocity peaks produced by
substructure against the background of random halo stars.  The main
reason to use RRLS as tracers is that even when they do not represent
the most numerous population of a system, they provide precise
distances and good velocities to help pin-point the location in the
sky where to continue searching for evidence of an accretion event. The
few RRLS and BHB stars one might find clustered in a particular place
in the sky are thus highly significant. The confusion caused by the use of a more
numerous tracer with a larger distance error in regions where several
stream candidates are suspected can be clarified by the use of RRLS. The
trade off is that the number of stars found will be smaller but their
parameters will be very reliable.
 
This paper is organized as follows: in Section~\ref{sec-sample} we
describe the sample of RRLS observed here and its relationship
(in space) with other works carried out in the
region. Section~\ref{sec-observations} explains the observations and
processing techniques that we have used. We look for coherent groups
in the sample of RRLS in Section~\ref{sec-results}, and investigate whether or not these groups are also found in surveys using different tracers.  Finally, we
discuss our results in Section~\ref{sec-conclusions}.

\section{THE SAMPLE OF RR LYRAE STARS \label{sec-sample}}

We have merged 3 different spectroscopic data sets: 18 RRLS from 
\citet{duffau06}, 
29 RRLS from \citet{vivas08}, and 36 RRLS which are reported here for the
first time.  In this latest data set most stars come from the QUEST survey \citep{vivas04} but
we also took spectra of 3 of the RRLS discovered by \citet{ivezic00}
\citep[and followed up by][]{wu05},  
with SDSS data just beyond the northern limit of QUEST.  
As we explain below, the radial velocity measurements of the sample RRLS
that was observed by \citet{duffau06} have been revised.
We have also updated the radial velocity measurements for some stars
in the sample of \citet{vivas08}, for which additional photometric and
spectroscopic data have been obtained.
For three of the stars in our sample we were able to obtain only their
mean metallicities, not their radial velocities. 

\begin{figure}
   \centering
     \includegraphics[angle=-90,width=8cm]{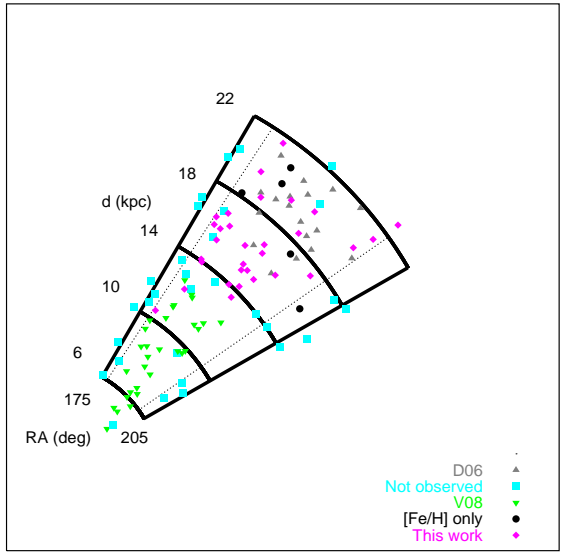}
      \caption{Distribution of the QUEST RRLS in the sky. The radial
        direction shows distance from the Sun (d), while the angular
        direction corresponds to right ascension (RA). Different symbols show the 3 data
        sets we are using in this work. It can be seen that the new
        observations (diamond symbols) fill the gap in distance
        between the samples of \citet[][D06]{duffau06} and
        \citet[][V08]{vivas08}. The square and circle symbols indicate
        those RRLS for which we did not obtain spectroscopic
        observations or we were not able to measure a radial
        velocity. We concentrate our analysis in the region between
        the two dotted radial lines ($178\degr <$ RA $< 200\degr$),
        in which we have phase space information for most of the QUEST
        RRLS.  }
         \label{fig-wedge}
\end{figure}

Most of our targets come from the photometric study performed by Vivas
and collaborators on two bands of QUEST data. Each band is a
$2.2\degr$-wide stripe at a fixed declination observed multiple times
in driftscan mode with the QUEST camera \citep{baltay02}, with the 1m
Schmidt telescope at the Venezuelan National Observatory of Llano del
Hato. The results of the survey in the band centered at
Dec $=-1\degr$ are reported in \citet{vivas04}; the second band, at
Dec $=-3\degr$, has not been published yet.  The selected
QUEST RRLS lie at $178\degr <$ RA $ < 200\degr$ and $-4.25\degr <
$ Dec $ < 0\degr$, while the 3 SDSS RRLS have the same range of
RA but $0\degr <$ Dec $< 1.23\degr$.  The RRLS have mean V
magnitudes between 13.5 and 17.5 mags, which correspond to a distance
from the Sun between 4 and 23 kpc. Figure~\ref{fig-wedge} is a polar
plot in which right ascension (RA) and distance from the Sun (d) for
our sample are shown in the angular
and radial direction, respectively. 

\begin{figure}
   \centering
   \includegraphics[bb=0 100 600 550, width=8cm]{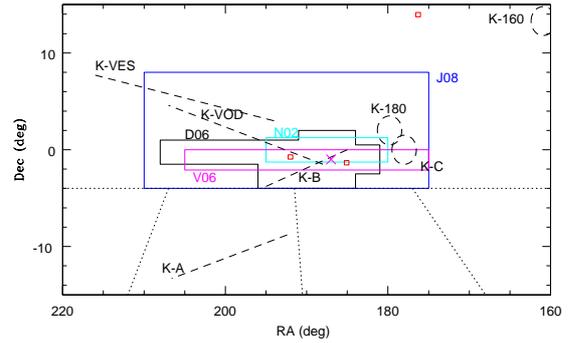}
      \caption{Distribution in the sky of the different photometric detections of sub-structures in the Virgo region.  
The dotted horizontal line at Dec $=-4\degr$ represents the rough
southern limit of the SDSS imaging survey.  
There are also 3 SEGUE imaging stripes that cross the region at negative declinations as shown by the other dotted lines.  
Labels correspond to the following references: N02, \citet{newberg02}; V06, \citet{vivas06}; D06, \citet{duffau06}; 
J08, \citet{juric08}; K-A, K-B and K-C, \citet{keller08,keller09}; K-VOD, K-VES, K-160 and K-180, \citet{keller10}. 
The small squares represent the fields observed by
\citet{jerjen13}; the one at (RA, Dec) $= (185\degr, -1\fdg 4)$
coincides with the location of Vir Z \citep{walsh09}. {The $\times$
symbol shows the densest part of the overdensity as seen in RRLS in the
QUEST survey}.}
         \label{fig-phot}
\end{figure}

\begin{figure}
   \centering
   \includegraphics[bb=0 100 600 550, width=8cm]{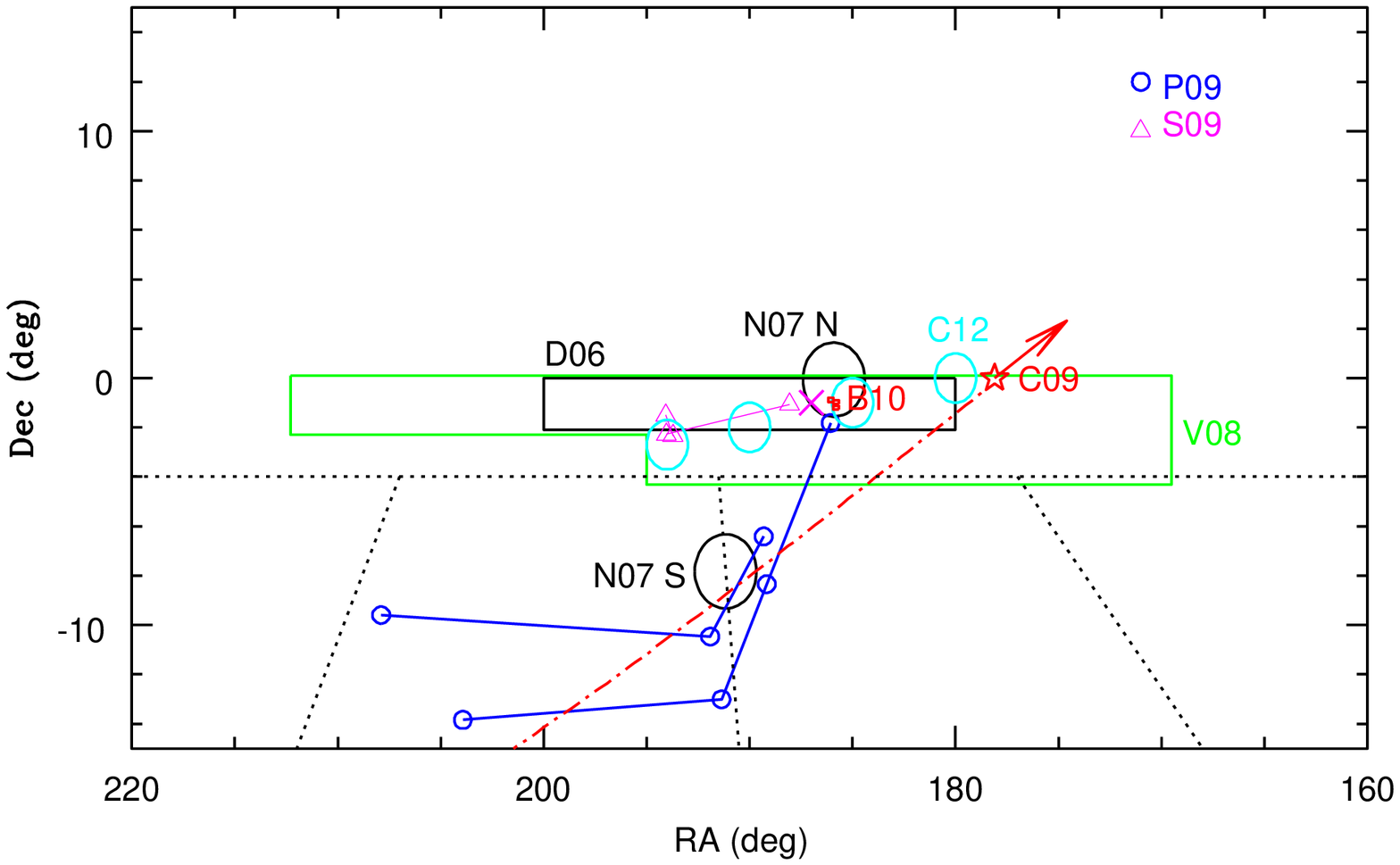}
      \caption{Similar to Figure~\ref{fig-phot} but for kinematic
        detections. The dotted lines and the $\times$ symbol are the
        same as in Figure~\ref{fig-phot}. Labels correspond to the
        following references: D06 (internal rectangle),
        \citet{duffau06}, N07 N and N07 S (large circles),
        \citet{newberg07}; V08 (polygon), \citet{vivas08}; S09 (small triangles),
        \citet{starkenburg09}; P09 (small circles), \citet{prior09a};
        B10 (tiny squares),
        \citet{brink10}; C12 (medium size circles), \citet{casey12}.  The star symbol, arrow
        and dot-dashed line correspond to the location of an RRLS whose
        proper motion was determined by \citet{casetti09} (C09), the
        direction of the proper motion and part of the past orbit,
        respectively.}
         \label{fig-spec}
\end{figure}

To compare our sample with previous ones, we show in
Figures~\ref{fig-phot} and \ref{fig-spec} the photometric and
spectroscopic detections of substructures in the region. In both
figures, the center of the densest part of the overdensity, as seen by the QUEST
survey, at (RA, Dec) =$(187\degr$,$-1\degr)$, is shown with a
$\times$ symbol.  Among the detections of substructures in
photometric data, there are overdensities of RRLS from the QUEST
\citep[V06]{vivas06} and SEKBO surveys \citep[groups K-A and K-B in
  Figure~\ref{fig-phot},][]{keller08}, F turnoff stars
\citep[N02]{newberg02}, main sequence stars in SDSS
\citep[J08]{juric08}, and sub-giant stars in SDSS
\citep[K10]{keller10}. In addition, overdensities
have appeared in the region by analyzing the luminosity function
 \citep{duffau06} and by combining main sequence and red giant
stars from the SEKBO survey \citep[K09]{keller09}. A strong main
sequence located at $\sim 23$ kpc was detected by \citet{jerjen13} in
the three fields shown in Figure~\ref{fig-phot}. One of those fields
is coincident with the location of an alleged faint satellite galaxy,
Vir Z, discovered by \citet{walsh09}. However, the results presented
by \citet{jerjen13} suggest that this object is more extended and it
is located much closer than initially thought. Except for the SEKBO survey,
which goes along the ecliptic, all other detections come from QUEST
and SDSS, which have a southern limit at about Dec $=-4\degr$ in this
part of the sky (shown as a horizontal dotted line in Figures~\ref{fig-phot} and
\ref{fig-spec}). Three SDSS/SEGUE stripes cross the region toward
southern declinations (also shown as dotted lines). From these stripes
and the SEKBO survey, it has been suggested that the VOD extends to
the south, beyond the SDSS and QUEST limits
\citep{newberg07,keller08,prior09a}.

Detections of substructure by radial velocity have been made through the measurement of either many stars in
relatively small fields, \citep{newberg07,brink10,casey12}, or 
relatively few stars but widely distributed over the region
\citep{duffau06,vivas08,prior09a,starkenburg09}. The latter usually
involves targets for which good distance determination is possible. Hence,
precise positioning of any stream in the sky is guaranteed, at the
expense of a low number of members. Following this idea,
\citet{duffau06} measured RRLS in the densest
part of the overdensity (as seen by QUEST) at 19 kpc, and found a
coherent velocity at $V_{\rm gsr} \approx 100$ ${\rm km~s}^{-1}$ (but see
revision below), which was dubbed the {\it Virgo Stellar Stream}
(VSS).  Also, the SEKBO survey of RRLS
\citep[P09][]{prior09a}, and red giants from the Spaghetti survey
\citep[S09][]{starkenburg09} found several groups of stars with
similar distance and radial velocity. Those groups are shown as
circles and triangles connected by lines in Figure~\ref{fig-spec}.

\citet{casetti09} measured the proper motion of one of the QUEST RRLS within the
region of the overdensity, which has a radial velocity consistent with being a member of the VSS. 
The measured star and the direction of the proper motion is shown 
with a star symbol and an arrow in Figure~\ref{fig-spec}. The path of a tentative past orbit
calculated by the authors is shown with a dot-dashed lined. This orbit has
been confirmed recently by \citet{carlin12} 
using proper motions for a more extended sample of turnoff stars.

\section{OBSERVATIONS \label{sec-observations}}

\subsection{Telescopes and setups}
 
 A total of 179 spectra of 82 RRLS (covering about 90 sq degrees in
 the sky), have been analyzed, including revisions of previous measurements.  The fainter stars were
 observed with the Magellan, WIYN, Blanco, and ESO 3.6m telescopes,
 while we used the 1.5m telescopes at both La Silla and CTIO (SMARTS)
 for the brighter ones.  Table~\ref{tab-tels} displays the details of all
 the telescopes and instrument setups used in this work.  Most of the
 SMARTS observations were obtained in service mode.

\begin{table*}
 \centering
\caption{Telescopes and Instruments}
\label{tab-tels}
\begin{tabular}{lccccccc}
\hline
\hline
Observatory & Telescope & Instrument & Code & Grating & Spectral range & Resolution & Dates \\
 &  &  & & ($mm^{-1}$) & (\AA) & (\AA) & \\
\hline
La Silla  &  1.5m  &  B\&C    & ESO1.5m & 600   & 3300-5500 & 3.1  &    June 2001 \\ 
La Silla &  3.6m  &  EFOSC2 & ESO3.6m & 600   & 3270-5240 & 5.4  & Jan-Feb 2006 \\
LCO      &  Clay   &  B\&C     &  Mag05  & 600   & 3110-6130 & 4.3  & March 2005 \\
LCO      &  Clay  &  LDSS3    &  Mag07   & 1090 & 3440-6140 & 2.1  & February 2007 \\
CTIO     &  4m    &  RC-BAS  & 4m          & 316   & 3480-6460 & 5.7 & Jan, Mar 2006 \\
CTIO     &  1.5m &  R-C        & SMARTS   & 600  & 3532-5300  & 4.3 & 2003-2004, 2006-2008 \\
KPNO    &  WIYN &  Hydra     & WIYN-B    & 400 & 3500-6200 & 7.1 & Apr 2006,Feb 2007,Feb 2008 \\
KPNO    & WIYN  & Hydra      & WIYN-R    & 600 & 7100-10000 & 2.9 & Mar 2003,Apr 2007,May 2007 \\ 
KPNO    & WIYN   & Hydra     & WIYN-G    &  600 & 4000-6800 & 4.6 &  May 2008  \\
\hline
\end{tabular}
\end{table*}

\begin{figure}
   \centering
   \includegraphics[angle=-90, width=8cm]{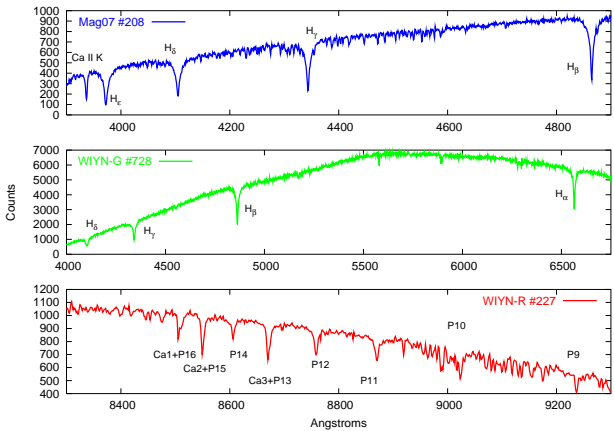}
      \caption{Example of reduced spectra of RRLS taken with some of the
      telescopes/setup combinations used for this project. The panels display
      the three wavelength ranges in use and identify typical spectral features in
      each of them. The telescope/setup labelling in each panel follows the code defined in
      Table~\ref{tab-tels}.}
         \label{fig-spectra}
\end{figure}

Most of the spectra were taken in the blue part of the optical window, allowing the
measurement of Balmer lines (excluding H$\alpha$) and the Ca II H and K lines. 
However, with the WIYN 
telescope we used also a ``red'' and a ``green'' configuration; this was done mainly to 
fit some of our targets within other scientific programs at this telescope.
In Figure~\ref{fig-spectra} we show the spectra of stars \#208, \#728 and \#227 which
represent spectra in the three different wavelength 
ranges covered by our telescope/setup combinations. Some characteristic spectral lines 
of RRLS are marked for reference. Only the blue spectra 
were used for measuring metallicities, which require a measurement of the CaII K line. Typical spectra for this project had a signal to noise of 20 
or larger, which is at the same time the minimum we used for estimating metallicities. 
Radial velocities could be estimated, if necessary, from spectra with a slightly lower signal to noise, but this happened in only a few cases since the quality of the majority of the spectra was high.

\subsection{Observing and Reduction Techniques}

Since RRLS are pulsating stars, the radial velocity changes during the
pulsation cycle.  In order to determine the systemic velocity of the
star, a curve must be fitted to the radial velocity
measurements. We tried to obtain at least two spectra per star, taken
at different phases during the cycle, but this was not always
possible.  As many as 4 spectra were obtained for a
few stars.

Since the radial velocity curve of a type ab RRLS has a large
discontinuity during the rising branch phase, we avoided taking
spectra at phases $>0.85$ and $<0.1$. Because most observations were
made in visitor mode, we were able to control the exact phase at which
each star was observed using the ephemerides from the QUEST
survey. Not only did we avoid the rising branch but we also spaced in
phase the observations for the same star for a better fitting of the
radial velocity curve.  In service mode, all the spectra were taken at
random phases. Some of them were not usable because they were obtained
on the rising branch, but since this is a rapid phase, only few
observations were lost. We did not control the phase of observation in
the WIYN/Hydra observations either, because several RRLS in a Hydra
field were observed at the same time.

When possible, we took comparison lamp exposures before and after each
program star, to account for any flexure of the instrument. In few cases,
spectra were reduced using only one of the lamps; in all others, the
average of the wavelength calibrations obtained with both lamps was
used to calibrate our spectra. The data were reduced using standard
IRAF\footnote{IRAF is 
distributed by the National Optical Astronomy Observatory, which is
operated by the Association
 of Universities for Research in Astronomy (AURA) under cooperative
 agreement
 with the National Science Foundation.} tasks or packages specifically made for an instrument. Wavelength
calibration was performed manually and carefully for each spectrum.
The wavelength calibrations had a typical RMS of 
$\sim 2$ ${\rm km~s}^{-1}$.

\subsection{New Photometry, New Ephemerides}

Following the methodology of \citet{vivas08}, we improved the
ephemerides of the QUEST RRLS by adding new photometric data for many
of our stars and by fitting an RRLS template to the light curves. The
new photometric observations came mainly from QUEST scans taken after
the publication of the first catalog.  A few additional observations were obtained with the 0.9 SMARTS telescope at CTIO. The fitting of the light curve used $\chi^2$ minimization
with either templates produced by \citet{layden98} for type ab stars,
or our own template for type c ones \citep{vivas08}.  We used the
original light curve parameters (namely, period, amplitude, phase at
maximum light and maximum magnitude) as initial values and allowed
them to vary.  The new observations of a few of the type c variables
showed that their published periods were spurious. The new light curve parameters
are part of the table with the final results of this paper
(\S~\ref{sec-results}).  The new photometry for these stars will be
included in an upcoming paper.

For most stars the changes in the light curve parameters were
minimal. But even a small change in the date at maximum light may have
an effect on the calculation of the phase of a spectrum.
There were a few stars ($\#$174, 180, 201, 203 and 217) for which the
changes were more significant. These stars are included in the new La
Silla-QUEST survey (\citet{zinn14}). Although these low amplitude variables were originally
classified as type c RRLS in the QUEST catalog \citet{vivas04}, the
light curves of La Silla-QUEST, which have a much larger number of
epochs ($\sim$80 to 190), have long periods and clear signs of
asymmetry. The shorter periods assigned by QUEST are aliases of the
true ones. These stars were then re-classified as type b and we used
the La Silla-QUEST light curve parameters in further calculations.

With these new improved ephemerides we re-calculated the phases at which our
old spectra from \citet{duffau06} and \citet{vivas08} were observed. 
Consequently, the final radial velocities for these data sets have changed and 
are updated here.

The La Silla-QUEST data also helped to identify star \#241 
\citep[reported as a type c in][]{vivas04,vivas08}
as a WUMa eclipsing binary. If the light curve is phased with twice the period reported in our 
previous work, it is easy to recognize that the two minima have different depths, a clear sign
of eclipsing systems. We therefore eliminated \#241 from further analysis.

\subsection{Radial Velocities and Metallicities}

\begin{table*}
\centering
\caption{Selected Layden stars for use as radial velocity and metallicity standards}
\label{tab-layden}
\begin{tabular}{lcc} 
\hline 
\hline
Star  & Radial Velocity & V \\ 
 & (${\rm km~s}^{-1}$) & \\ \hline 
Feige 56  & 34.6 & 11.1 \\ 
Kopff 27  & 5.5 & 10.2 \\
HD 693  & 17.0 & 4.9  \\
HD 22413  & 34.8 & 8.7  \\
HD 65925  & -8.2 & 5.2  \\
HD 74000 & 205.6 & 9.4 \\ 
HD 74438  & 20.4 & 7.9  \\
HD 76483  & 5.4 & 4.9   \\
HD 78791  & 22.4 & 4.5 \\ 
HD 140283  & -171.4 & 7.3 \\
HD 154417  & -18.0 & 5.8 \\ \hline
\end{tabular}
\end{table*}

\begin{table*}
\centering
\caption{Selected Preston-Sneden stars for use as radial velocity standards}
\label{tab-preston}
\begin{tabular}{lccccccc} 
\hline
\hline
Star & RA & Dec & V & (B-V) & [Fe/H] & Radial Velocity \\ 
\hline
 & (J2000) & (J2000) &  &  &  & (${\rm km~s}^{-1}$) \\
\hline
22175-034   & 02 20 21.4 & -10 38 10  & 12.6 & 0.34 & -0.28 & 27.2 \\
22185-009  & 03 14 53.8 & -14 43 46   & 13.8 & 0.30 & -1.67 & -100.6 \\
22874-009   & 14 34 23.0 & -26 17 39   & 13.7 & 0.24 & -0.42 & -36.7 \\
22874-042   & 14 38 01.7 & -24 58 47   & 14.0 & 0.33 & -1.53 & 176.2 \\
\hline
\end{tabular}
\end{table*} 

\begin{table*}
\centering
\caption{Selected ELODIE stars for use as radial velocity standards}
\label{tab-elodie}
\begin{tabular}{lcccccc} 
\hline 
\hline
Star & RA & Dec & B & V & Spectral Type & Radial Velocity \\ 
\hline
 & (J2000) & (J2000) &  &  &  & (${\rm km~s}^{-1}$) \\
\hline
HD 102870  & 11 50 41.7185 & +01 45 52.985 & 4.16 & 3.61 & F9V & 4.43 \\ 
HD 13555   & 02 12 48.0855 & +21 12 39.575 & 5.64 & 5.24 & F5V & 5.55 \\
HD 136202  & 15 19 18.7977 & +01 45 55.468   & 5.60 & 5.10 & F8III-IV & 54.41 \\
HD 140283  & 15 43 03.0966 & -10 56 00.590   & 7.69 & 7.24 & sdF3 & -169.83 \\
HD 16895   & 02 44 11.9863 & +49 13 42.412   & 4.60 & 4.12 & F7V & 24.44 \\
HD 19994   & 03 12 46.4365 & -01 11 45.964   & 5.63 & 5.06 & F8V & 19.21 \\
HD 222368   & 23 39 57.0409 & +05 37 34.650   & 4.64 & 4.13 & F7V & 5.62 \\
HD 22484   & 03 36 52.3832 & +00 24 05.982   & 4.86 & 4.28 & F9IV-V & 28.08 \\
HD 22879   & 03 40 22.0645 & -03 13 01.133   & 7.23 & 6.74 & F9V & 120.19 \\
HD 3268   & 00 35 54.8015 & +13 12 25.442   & 6.93 & 6.41 & F7V & -23.40 \\
HD 49933   & 06 50 49.8319 & -00 32 27.175   & 6.14 & 5.78 & F2V & -12.57 \\
HD 693   & 00 11 15.8573 & -15 28 04.719   & 5.38 & 4.89 & F8Vfe-08H-05 & 15.04 \\
HD 7476   & 01 14 49.1720 & -00 58 25.661   & 6.12 & 5.70 & F5V & 25.49 \\
\hline
\end{tabular}
\end{table*}

  Except for the case of the red setup for WIYN data, the radial
  velocities of the target stars were measured with the IRAF task {\it
    FXCOR}.  {\it FXCOR} performs fourier cross-correlation between a
  program star and a radial velocity standard star. In the case of the
  redder WIYN data, we calculated the velocities in a different way,
  by measuring the centers of some spectral lines. In both cases we
  followed very closely the methods explained in \citet{vivas08},
  which will not be repeated here. We discuss only relevant changes in
  the strategy applied specially to the data of the largest telescopes
  (Magellan and the 3.6m telescope).

The cross-correlation technique requires selecting radial velocity
standards of similar spectral types as the target to be observed
during the corresponding run. For all the runs, we observed several of
the radial velocity standard stars from \citet{layden94} (from now on
simply Layden standards). They are listed in Table~\ref{tab-layden}.
Velocities in this table are updated values taken from the SIMBAD
database.  Not all the standards in Table~\ref{tab-layden} were
observed in every run, but we observed several each night. Some
standards were observed several times during the same night or in
different nights within the same observing run.  The Layden
standards have the advantage of serving as metallicity standards as
well.  We used these standards for cross-correlating all SMARTS and
ESO 1.5m observations.  However, they are bright stars which are hard
to observe at large telescopes.  During the progress of this work we
found the need to set up different standard stars.  One set came out
of a group of Blue Metal Poor stars studied by \citet{preston00}
(from now on, Preston-Sneden standards). These standards were observed
only in the Mag07 run.  The Preston-Sneden standards are closer in magnitude to our target stars. We selected four non
variable stars from \citet{preston00} which were available in the sky
at the time of the observing run.  From \citet{preston00}, these stars
have small dispersions in their radial velocities, from 0.52 ${\rm
  km~s}^{-1}$ to 0.73 ${\rm km~s}^{-1}$.
 
Finally, we created another set of standards by degrading high
resolution spectra obtained from the ELODIE archive \citep{moultaka04}
to the resolution of our spectra. We did this for each of the
telescope-instrument setups.  We refer to this set as the ELODIE
standards. These stars have velocity estimates\footnote{private communication, 
Sergio Ilovaisky (Observatoire de Haute-Provence).}
 better than 0.1 ${\rm
  km~s}^{-1}$. Tables~\ref{tab-preston} and
\ref{tab-elodie} contain all of the relevant information of the radial
velocity standard stars used in these two sets.

For the ELODIE standard set we selected 13 stars with spectral types
F, which are similar to the ones of RRLS.  To degrade the
ELODIE spectra to the proper resolution we used the IRAF tasks {\it
  gauss}.  These degraded spectra were then binned to match the
binning of the targets using {\it dispcor}. Finally, we normalized the
spectra by using {\it continuum}, and trimmed the large spectral range
of the ELODIE spectra to match that of the target spectra.

Errors were estimated for each set of standard stars and instrumental
setup by cross correlating every standard with all others, and getting
the difference between the velocity obtained for each star and its
velocity from the literature.  This procedure revealed that some of
our observations of Layden standard stars were off.  Centering these
bright stars incorrectly on the slit may have produced a significant
shift in the spectrum.

For the Mag07 run we used the Preston-Sneden standards for
cross-correlation, and we used the ELODIE standards for the 3.6m
telescope observations. We then decided to revise the measurements of
the stars from the Mag05 run, which were reported in
\citet{duffau06}. The use of ELODIE standards on these observations
produced significantly better results, and the new measurements agree
with the results obtained by \citet{newberg07} for F turnoff stars in
the region (see \S~\ref{sec-revDuf}).

The estimated error for the radial velocities ($\sigma_r$) of the
Mag07 run is 13 ${\rm km~s}^{-1}$, while for the 3.6m telescope
is 9 ${\rm km~s}^{-1}$.  The estimated errors from the SMARTS, WIYN
(blue, green and red setups) and ESO 1.5m telescope data sets are, 16,
8, 6, 9 and 12 ${\rm km~s}^{-1}$, respectively.
 
Unfortunately, the data from the 4m Blanco telescope at CTIO 
proved to be unsuitable for radial velocity measurements, as shifts of
one pixel or more were found in the raw data on consecutive exposures
of the same star. Nevertheless, we could measure the metal abundances of these stars
because the typical shifts (of about 1 pixel), represent $\sim 1$ \AA, which is
very small compared the width of the spectral features used in Layden's method of determining
[Fe/H].  Using this method, which is only applicable to the blue
spectra, we measured [Fe/H] with precisions of 0.15 and 0.2 dex for
types ab and c, respectively.  See our previous investigations for
more details \citep{vivas05,duffau06,vivas08}.

Although the dependence of the absolute magnitude of RRLS ($M_V (RR)$)
on metallicity is small, we took advantage of the fact that we had spectroscopic
measurements for almost all stars in our sample.  We calculated the
distances to the stars using the relationship in \citet{demarque00}:

\begin{equation}
\centering
M_V (RR) = 0.22\mbox{[Fe/H]} + 0.90
\label{eq-mvfe}
\end{equation}

For the few stars for which we did not measure a metallicity, we assumed $M_V (RR) = +0.55$,
which corresponds to a metallicity of [Fe/H]$=-1.6$ in the above equation.

\section{Results \label{sec-results}}

Table~\ref{tab-obs} contains all the individual observations of the RRLS 
investigated here. Many stars have multiple entries in this table since they
were observed more than once at different epochs.
The table includes the revised velocities of the stars
first reported in \citet{duffau06}. 
The phase at which the spectra in \citet{duffau06} and \citet{vivas08}
were taken has been recalculated in this table using the new ephemerides. 

\begin{table*}
\centering
\caption{Individual observations of the target RR Lyrae stars}
\label{tab-obs}
\begin{tabular}{lccccccc} 
\hline
\hline
Star & HJD & Exp & Telescope & Phase & $V_r$ & $\sigma_r$ & [Fe/H] \\ 
\hline
 & (+2450000 d) & (s) &  &  & (${\rm km~s}^{-1}$) & (${\rm km~s}^{-1}$)  &  \\
\hline
172 &  3737.8304 &  520  & ESO3.6m  & 0.270 &  -25 &  11 & -1.75 \\
175 &  3831.6760 & 1200 & WIYN-B & 0.208 &  180 &  15 & -2.19 \\
175 &  4153.8770 & 3600 & WIYN-B & 0.572 &  292 &   8 & -2.36 \\
176 &  3737.8091 &  400  & ESO3.6m  & 0.538 &  396 &  15 & -1.41 \\
176 &  4153.8770 & 3600 & WIYN-B & 0.023 &  366 &   8 & ... \\
177 &  3436.6576 &  800  &  Mag05 & 0.565 &  266 &  17 & -2.26 \\
\hline 
\end{tabular}
\tablefoot{This table is available in its entirety in the only version
  of this article. A portion is shown here for guidance 
regarding its form and content. The table displays the ID of the star,
following the QUEST catalog, the heliocentric julian date for the
observation, the exposure time in seconds, the telescope/setup used,
the estimates for phase, radial velocity, radial velocity error, and metallicity. The telescope/setup column corresponds to the codes defined 
in Table~\ref{tab-tels}.}
\end{table*}

\begin{figure}
\centering
\includegraphics[width=8cm]{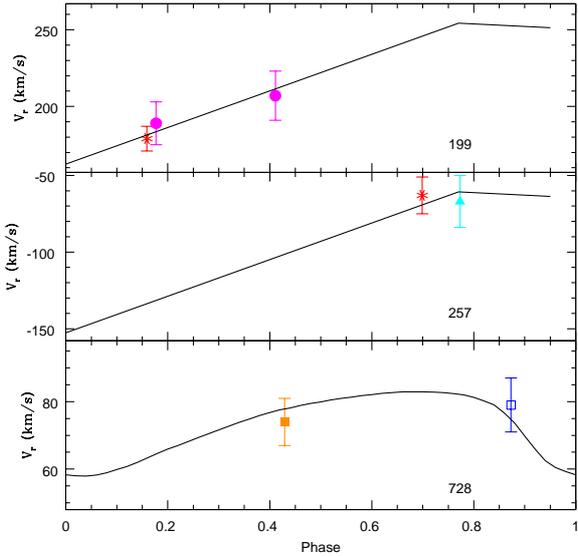}
\caption{Example of radial velocity curves fitted to the data of stars
  199, 257 and 728. The first two are RRab while the last one (bottom
  panel) is a type c star. These stars were observed with different
  telescopes. The consistency of the data is reassuring that no large
  systematic offsets are present in the data. Solid circles represent
  velocities measured in Magellan runs; asterisks, open square and
  solid square correspond to different WIYN setups (red, blue and
  green, respectively); triangle is a measurement from the ESO 3.6m
  telescope. The complete set of radial velocity figures are available as online only material.}
\label{fig-rv}
\end{figure}

Figure~\ref{fig-rv} shows examples of
the radial velocity curves fitted to three different stars. The
complete sample of radial velocity curves for stars having 2 or more
observations are available in the online version of this article.  As can be seen in
Figure~\ref{fig-rv} (and its online companions), there is good
consistency among the results from different instrumental setups,
which give us confidence that there are no significant systematic
differences among the data sets.  Within errors, there is also
consistency in the metallicities determined for the same star from
data coming from different telescopes.  Table~\ref{tab-vel} contains
the final results for all the 82 stars in the sample, after fitting the
radial velocity curves and averaging metallicities obtained from the
individual spectra.  Because of the
revision on velocities and ephemerides, this table supersedes the one
in \citet{vivas08}.  Table~\ref{tab-vel} contains: ID, RA, Dec, mean
extinction-corrected magnitude (V$_0$), period in days, heliocentric
Julian day at maximum light (HJD$_0$), type of RRLS, distance from the
Sun in kpc (d), number of spectra used for the radial velocity curve
fitting (Nfit), heliocentric radial velocity ($V_\gamma$), rms of the
fitting of the radial velocity curve ($\sigma_{fit}$), final error in
the systemic velocity ($\sigma_\gamma$), radial velocity in the
galactic standard of rest (V$_{\rm gsr}$), and metallicity [Fe/H]. For
details on how the final error in the systemic velocities were
calculated we refer the reader to \citet{vivas08}.

\begin{table*}
\centering
\scriptsize
\caption{RR Lyrae data: positions, light curve parameters, systemic velocities and metallicity}
\label{tab-vel}
\begin{tabular}{lccccccccccccc} 
\hline
\hline
ID & RA & Dec & V$_0$ & Period & HJD$_0$ & Type & d & Nfit & $V_\gamma$ &
$\sigma_{fit}$ & $\sigma_\gamma$ & $V_{\rm gsr}$ & [Fe/H] \\
\hline
 & (deg) & (deg) & & (d) & (+2450000 d) &  & (kpc)  &  & (${\rm km~s}^{-1}$) & (${\rm km~s}^{-1}$) & (${\rm km~s}^{-1}$) & (${\rm km~s}^{-1}$)  \\
\hline
172 & 180.185025 & -0.712449 & 16.54 & 0.52119 & 1610.71317 & ab & 16.0 & 1 & 2 & 0 & 17 & -104 & -1.75 \\
175 & 180.632505 & -1.895691 & 16.14 & 0.57091 & 1582.74254 & ab & 14.1 & 2 & 249 & 34 & 34 & 140 & -2.28 \\
176 & 181.087335 & -2.178318 & 16.38 & 0.33061 & 1971.84326 & c & 14.4 & 2 & 385 & 4 & 7 & 276 & -1.41 \\
177 & 181.212600 & -0.351735 & 17.03 & 0.37668 & 1585.81581 & c & 21.0 & 2 & 261 & 3 & 13 & 158 & -2.19 \\
182 & 182.130810 & -0.788163 & 16.80 & 0.59315 & 1620.71096 & ab & 18.8 & 1 & -25 & 0 & 22 & -128 & -2.16 \\
\hline 
\end{tabular}
\tablefoot{This table is available in its entirety in the online
  version of this article. A portion is shown here for guidance 
regarding its form and content. See text for description.}
\end{table*}

\subsection{Revision of \citet{duffau06} \label{sec-revDuf}}

 From the analysis of our first spectroscopic data (18 RRLS and 10 BHB
 stars from \citet{sirko04a}), we concluded in \citet{duffau06} that
 there was kinematical evidence for a stellar stream (the VSS) at
 $\sim 19$ kpc from the Sun.  These first results indicated that a
 velocity peak at $V_{\rm gsr}$ $\sim$ 100 ${\rm km~s}^{-1}$ with a
 dispersion of $\sigma \sim$ 17 ${\rm km~s}^{-1}$, and a mean metal
 abundance estimated in [Fe/H] = -1.86 and $\sigma$ = 0.40 existed
 within the clump of RRLS.  The BHB stars in that work were taken from
 SDSS DR1, as selected by \citet{sirko04a}.  We use below the updated velocities
 of these BHB stars from SDSS DR7.

\begin{figure}
   \centering
   \includegraphics[bb=0 320 600 720, width=8cm]{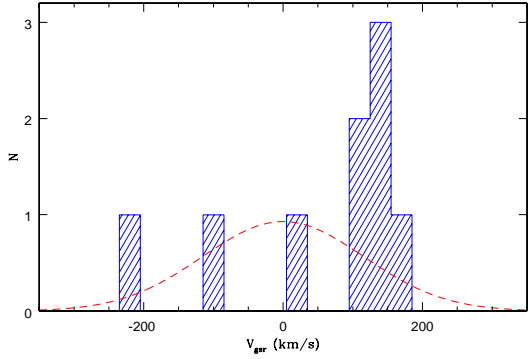}
      \caption{Radial velocity distribution of 9 RRLS and 1 BHB star
        in the densest part of the VSS region. For reference, the
        expected distribution of the same number of halo stars at the
        distance of this region is shown as a Gaussian curve
        ($\langle{V_{\rm gsr}}\rangle = 0$ ${\rm km~s}^{-1}$,
        $\sigma=116$ ${\rm km~s}^{-1}$).  }
         \label{fig-revDuf}
\end{figure}

Figure~\ref{fig-revDuf} displays the new velocity histograms for the
densest region of the overdensity (at $\sim 19$ kpc), as defined in
\citet{duffau06}. There are 9 stars (8 RRLS and 1 BHB star) in this
region.  In the revised histogram shown in Figure~\ref{fig-revDuf},
which can be directly compared with Fig. 2 (top) in \citet{duffau06},
there is a clear peak in the velocity distribution, which
contains the same 6 stars as in the previous paper, but the peak is
slightly shifted toward more positive velocities. The new mean of this
kinematical group is $V_{\rm gsr} = 128$ ${\rm km~s}^{-1}$ and it has a
standard deviation of $\sigma= 19$ ${\rm km~s}^{-1}$.  The mean
velocity is now even further away from the model halo without
rotation, (0 ${\rm km~s}^{-1}$), and the standard deviation is still
only slightly higher than our mean observational error.  More
importantly, the new mean velocity now agrees very well with the result of
$V_{\rm gsr}=130$ ${\rm km~s}^{-1}$ that \citet{newberg07} found in the
S297+63-20.5 feature from spectroscopy of turnoff F stars. Although
S297+63-20.5 and the VSS are virtually in the same part of the sky
(see Figure~\ref{fig-phot}) and at the same distance,
\citet{newberg07} were reluctant to associate both features because of
the offset in velocity. The revision of our velocities, removes the
offset to within observational errors.  Thus, the VSS and S297+63-20.5
refer to the same halo substructure.

\subsection{Radial Velocity Distribution in the Virgo Region}

\begin{figure}
   \centering
   \includegraphics[width=8cm]{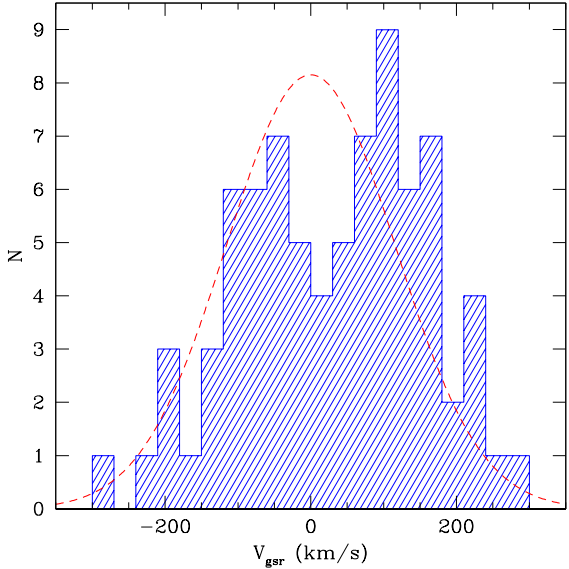}
      \caption{Radial velocity distribution of 79 RRLS in the Virgo
        region. Bin size is 30 ${\rm km~s}^{-1}$. For comparison, the
        expected distribution of the same number of halo stars at the average distance of
        these RRLS is shown as a Gaussian curve ($\langle{V_{\rm
            gsr}}\rangle = 0$ ${\rm km~s}^{-1}$, $\sigma=116$ ${\rm
          km~s}^{-1}$).  }
         \label{fig-histogram}
\end{figure}

Figure~\ref{fig-histogram} shows the distribution of radial velocities
for all the stars in the sample. We proceed to examine the velocity histogram for 
deviations with respect to the expected smooth non-rotating
  halo model. The mean radial velocity is 31 ${\rm
  km~s}^{-1}$ with a standard deviation of 124 ${\rm km~s}^{-1}$.  The
expected distribution of halo field stars at the mean distance of the
sample of these RRLS, $\sim 15$ kpc, has a standard deviation of $\sim
116$ ${\rm km~s}^{-1}$, according to the velocity profile measured by
\citet{brown10}, and it is shown by the dashed Gaussian line in
Figure~\ref{fig-histogram}. From this figure it is clear that there
are important deviations of our data with respect to the expected
normal distribution, which suggest that 
substructures may be present. Because we have taken the complete sample of stars to 
make this histogram, the information about the distribution in distance of the velocity deviations 
is not evident.

\begin{figure}
   \centering
   \includegraphics[width=8cm]{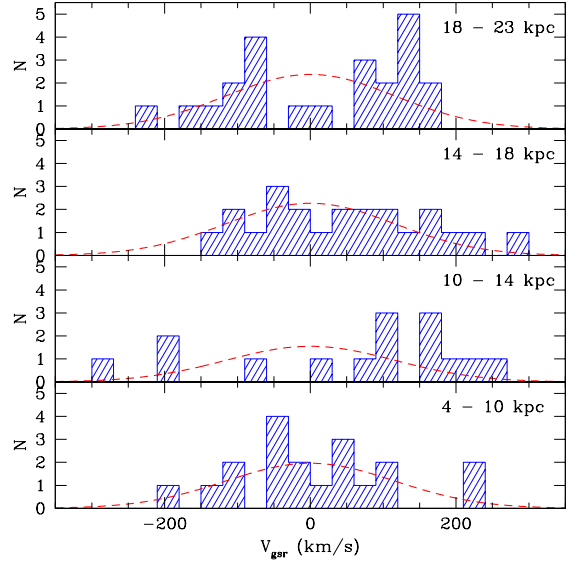}
      \caption{Radial velocity distribution of our stars in 4 different distance
        bins. Gaussian curves representing a halo smooth distribution
        are shown for the same number of stars contained in each
        distance bin. }
         \label{fig-hbins}
\end{figure}

A better sense of where the kinematic groups may be located can be seen by splitting the
velocity histogram into four distance bins (Figure~\ref{fig-hbins}).
We examined each bin separately, 
and found evidence of velocity groups in two of them.

{\bf 18-23 kpc bin:} The most distant bin (top panel in Figure~\ref{fig-hbins}) shows the
distribution of the 23 RRLS with $d>18$ kpc.  The distribution of
velocities is definitely non-Gaussian and this is confirmed by a
Shapiro-Wilk test which rejects normality at the 97\% confidence
level.  This panel corresponds to the range of distances where the VSS
was first detected by \citet{duffau06}. The VSS is seen in this
histogram as 7 stars with $120 \lessapprox V_{\rm gsr} \lessapprox
180$ ${\rm km~s}^{-1}$. But, with the inclusion of more stars to the
original \citeauthor{duffau06}'s sample, a second peak in the
distribution is now easily observed at $V_{\rm gsr} \sim -75$ ${\rm
  km~s}^{-1}$.

{\bf 10-14 kpc bin:} The distribution of velocities in the range of distances $10 < d < 14$
kpc also shows hints of substructures since an excess of stars at very
high positive velocity ($> 150$ ${\rm km~s}^{-1}$) is present.  This
peak is mostly made out of the stars already reported in
\citet{vivas08}.  A second peak is seen at $V_{\rm gsr} = 105$ ${\rm
  km~s}^{-1}$.  In this distance bin, normality in the velocity distribution is
also rejected by the Shapiro-Wilk test at the 95\% confidence level. The
stars in this bin have a mean velocity $\langle{V_{\rm
    gsr}}\rangle = 57$ ${\rm km~s}^{-1}$, which is significantly
offset from the nominal value of the halo; this is of course due to
the presence of the two positive velocity groups.  The standard deviation
is also significantly larger than the nominal halo, $\sigma=166$ ${\rm
  km~s}^{-1}$.

{\bf bins 4-10 kpc and 14-18 kpc:} These two bins do not show 
significant features in the velocity
distribution.  According to Shapiro-Wilk statistical test there is no
reason to suspect the stars in these bins are not drawn from a normal
distribution. 

Since we chose arbitrarily the limits of the distance bins in the four
  panels of Figure~\ref{fig-hbins}, it is possible that the
  identified velocity groups extend beyond the limits where they are recognized.  In the following sections, we will
  apply a group finding algorithm to the RRLS without imposing distance bins.

\subsection{Group Finding Algorithm \label{sec-groups}}

To find groups of stars in phase space we have used the idea of {\sl
  Stellar Pairs}, inspired by the works of
\citet{clewley06,vivas08} and \citet{starkenburg09}. Two stars make a
stellar pair when they are close in space and their velocities are
similar. A group is formed when two or more pairs have stars in
common. Thus, the minimum number of members in a group is 3 (two pairs
with 1 star in common).
 
\citet{starkenburg09} define a "pair" of stars based on the {\sl four-distance}, $4d_{ij}$, between stars 
$i$ and $j$:

\begin{equation}
\centering
(4d_{ij})^2 = \omega_\phi^2 \phi_{ij}^2 + \omega_d (d_i-d_j)^2 + \omega_v (v_i-v_j)^2 
\end{equation}

\noindent
where the first term on the right hand of the equation represents the
angular distance of two stars in the sky, and the second and third
terms are their differences in distance from the Sun and radial
velocity, respectively. Each term is weighted by the values
$\omega_\phi$, $\omega_d$ and $\omega_v$, defined in
\citet{starkenburg09}, to take into account the observational errors
and to normalize to the maximum possible separation between the two
stars.  With this definition, two stars make a pair when $4d_{ij} <
\epsilon$, where $\epsilon$ is a free parameter that gives a measurement of how
tight the groups are in phase space.

A potential problem with the above formalization is that the maximum allowed separation in the sky (measured in degrees) means different physical 
scales of the projected size in kpc of a group of stars in the sky, 
depending on their distance from the Sun. This means
that with increasing distance, the method imposes the search of groups which are tighter in the
direction perpendicular to the line of sight compared with the allowed size along the
line of sight. This is not appropriate for the search of cold streams,
where the constraint in distance separation must remain the same along the stream
independent of orientation or distance from the Sun.

Here we propose a modification on the definition of the four-distance to ensure that at any
place in the Galactic halo, the method does not impose any preference to the physical
scale along one dimension or the other:

\begin{equation}
\centering
(4d_{ij})^2 = \omega_{3d_{ij}}^2 (3d_{ij})^2 + \omega_v (v_i-v_j)^2 
\label{eq-new4d}
\end{equation}

\noindent
where $3d_{ij}$ is the three-dimensional distance between two stars calculated
from their galactocentric cartesian coordinates (x,y,z):

\begin{equation}
\centering
3d_{ij} = \sqrt{(x_i-x_j)^2 + (y_i-y_j)^2 + (z_i-z_j)^2}
\end{equation}

The weights in equation~\ref{eq-new4d}, $\omega_{3d_{ij}}$ and $\omega_v$, are defined following 
\citet{starkenburg09} as:

\begin{equation}
\omega_{3d_{ij}} = \left( \frac{1}{(3d)_{max}} \right) ^2 \frac{\left( \frac{\sigma d_i}{d_i} \right)^2 +
\left( \frac{\sigma d_j}{d_j} \right)^2}{2 \langle \frac{\sigma d}{d} \rangle ^2} 
\end{equation}

\begin{equation}
\omega_v = \left( \frac{1}{v_{max}} \right) ^2 \frac{\sigma v_i^2 + \sigma v_j^2}
{2 \langle \sigma v \rangle ^2}
\end{equation}

The weights are normalized by the maximum possible range in distance and velocity for two stars 
which we set
to $(3d)_{max}=60$ kpc and $v_{max}=550$ ${\rm km~s}^{-1}$. With a
$\sigma$ of 116 ${\rm km~s}^{-1}$ \citep{brown10} for the distribution
of the radial velocities of halo stars, the value of $v_{max}$
encloses $2.2 \sigma$ of the distribution. On the other hand, the
number of RRLS farther away than 60 kpc from the galactic center is
very low \citep{sesar07,zinn14} and thus it is an appropriate value
for $(3d)_{max}$. For the weight in the (3d) 
distance we only took into account  the error in distance along the
line of sight since the errors in position of the stars (less than
$1\arcsec$) are negligible compared with the typical separation
between stars in a pair (a few degrees).

The critical value $\epsilon$ establishes how small is the
separation of two stars in space and in velocity. It is a measure of the scale of
clumpiness of the groups. The smaller the values of $\epsilon$, the
tighter the groups of stars. Table~\ref{tab-epsilon} shows the maximum values
in 3d distance and velocity for stars in a pair for different values of $\epsilon$, which
are obtained by making two stars having one of the parameters (either 3d distance or velocity)
exactly the same. In practice stars in a pair will have separations smaller than the values 
quoted in the table since it is unlikely to have two stars with exactly the same position or velocity. 
Small values of $\epsilon$ may impose
constraints which are too tight for the expected density of stars in a
stream, velocity dispersion of a typical stream, and for the 
observational errors of our sample. For example, the maximum separation in velocity of
a pair of stars with $\epsilon=0.02$ is only 10 ${\rm km~s}^{-1}$, which is smaller than
our observational errors. On the other hand, very large values of
$\epsilon$ may be unable to distinguish individual groups, 
since they will pair the whole sample into one big group.

\begin{table}
\caption{Maximum values for stars in a pair}
\label{tab-epsilon}
\begin{tabular}{ccc} 
\hline 
\hline
$\epsilon$ & 3d (kpc) & v (${\rm km~s}^{-1}$)  \\ 
\hline 
0.02 & 1.2 & 10 \\
0.03 & 1.8 & 15 \\
0.04 & 2.4 & 20 \\
0.05 & 3.0 & 25 \\
0.06 & 3.6 & 30 \\
\hline
\end{tabular}
\end{table}

\begin{figure}
   \centering
   \includegraphics[width=8cm]{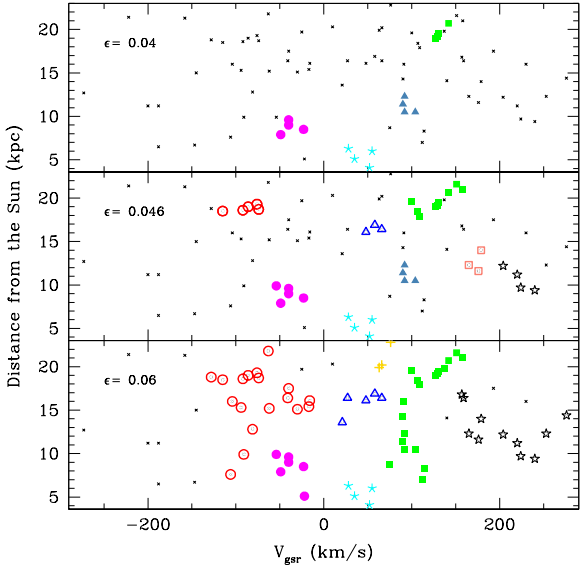}
      \caption{Distance from the Sun versus radial velocity for all the RRLS in the sample.
      Groups detected with $\epsilon = $ 0.04 (top), 0.046 (middle) and 0.06 (bottom) 
      are shown with different colors/symbols. The properties of the groups shown in 
      the middle panel ($\epsilon=0.046$) are specified in Table~\ref{tab-groups}.}
         \label{fig-epsilon}
\end{figure}

For our RRLS sample we explored
different values of $\epsilon$ in the range from $0.02$ to $0.07$ and we
found pairs and groups with all of them. We show in Figure~\ref{fig-epsilon} some
examples of the groups we found with three different values of $\epsilon$ 
(0.04, 0.046 and 0.06). The plots show the distribution in phase space of each one of the groups
that were detected with a different color/symbol.
The same group may be detected with different values of $\epsilon$.
A tight group (small $\epsilon$) may get more members with increasing $\epsilon$
(for example, the solid green squares at $\sim 130$ ${\rm km~s}^{-1}$),
unless it is physically a very concentrated and isolated clump in
phase space
(for example, the solid cyan stars at $\sim 40$ ${\rm km~s}^{-1}$). 
Other groups just appeared for the first time with the largest value of $\epsilon$.
In the figure it can be seen that both the number of groups and the number of 
members in each group
increase with $\epsilon$, until $\epsilon=0.06$ where the number of groups starts
to decline. This is a sign that the maximum allowed separations in distance and
velocity are too large and
several groups have merged together. 
In addition, for $\epsilon \ge 0.05$ we obtained several groups that
have velocity dispersions larger than the expectations for cold substructures in the
halo (which should be below $\sim 30$ ${\rm km~s}^{-1}$). Some groups also have
an unusually large range in distance along the line of sight, covering in some cases almost
the whole range of the RRLS sample. Examples of this can be seen in the lower panel
of Figure~\ref{fig-epsilon} which shows the results for $\epsilon=0.06$. Some of the
groups (for example, the red open circles at $\sim -75$ ${\rm km~s}^{-1}$) are just
too dispersed both in distance and velocity to believe they are part of a real cold substructure. 
Thus, the optimal value of $\epsilon$ for finding cold substructures in the RRLS samples 
seems to be a value below 0.05.
 
We found that $\epsilon=0.046$ allowed for the loosest constraints that
produced the maximum number of members in each group but
still having velocity and distance dispersions expected for halo substructures. 
In the following we study with detail the groups found with this value of $\epsilon$.
Table~\ref{tab-groups} indicates the mean properties of the 8 groups (labelled with
letters A to H) detected
with this value of $\epsilon$, including mean distance and velocity, velocity dispersion, distance
dispersion (along the line of sight)
and number of members. In addition, 
we calculated the 3d dispersion, $\sigma 3d$, of each group by obtaining
the coordinates of the center of the group from the average of 
the cartesian coordinates of the members. Then,
we estimated the distance separation of each member to the center of the group. The
3d dispersion, which is given in kpc, is the average of those distance separations.
Finally the table contains some statistics that help to determine if the group
is a significant feature of the halo, as we explain in the next
section.

\begin{table*}
\centering
\caption{Properties of the Groups of RRLS in the Virgo Region detected with $\epsilon=0.046$}
\label{tab-groups}
\begin{tabular}{cccccccccccc} 
\hline 
\hline
Group  & Color/ & N & $\langle V_{gsr} \rangle$ 
& $\sigma V_{gsr}$ & $\langle d \rangle$ & $\sigma d$ & $\sigma 3d$ & 
P($V_{gsr}$) & P(d) & P($\sigma d$) & P($\sigma 3d$) \\ 
  & Symbol &  &  (${\rm km~s}^{-1}$)  & (${\rm km~s}^{-1}$)  & (kpc)  & (kpc) & (kpc)  &  &  &  & \\ 
\hline
A & green/solid squares   & 9 & 128 & 18.9 & 19.7 & 1.15 & 1.57 & 99.6 & 98.8 & 11.3  & 28.7 \\
B & red/open circles         & 5 & -89 & 14.7 & 18.8 & 0.29 & 1.42 & 79.2 & 90.9 & 0.6 & 58.1 \\
C & magenta/solid circles & 5 & -41 & 10.6 &   9.0 & 0.72 & 0.95 & 43.3 & 37.7 & 15.4  & 13.3 \\
D & cyan/solid stars                  & 4 &   43 & 11.3 &   5.4 & 0.86 & 0.81 & 39.4 & 3.2 & 40.3 & 12.1 \\ 
E  & blue/solid triangles    & 4 &  95  &   5.5 &  11.2 & 0.75 & 0.80 & 77.0 & 50.4 & 29.8 & 11.6 \\
F  & black/open stars		   & 4 & 222 & 12.8 & 10.6 & 1.14  & 1.45 & 99.8 & 46.4 & 66.0 & 75.3 \\
G  & blue/open triangles   & 3 & 57   & 7.4   & 16.5  & 0.33 & 1.05 & 46.3 & 72.4 & 11.4 & 52.0 \\
H  & pink/open squares & 3 & 173 & 6.0 & 12.6 & 1.01 & 1.11 & 96.4 & 51.9 & 72.7 & 59.3 \\
\hline
\end{tabular}
\tablefoot{N is the number of members of the indicated group,
  P(variable) is the percentile for that group in that particular
  variable. See text for description.}
\end{table*}

\subsection{Simulations\label{sec-sim}}

A smooth distribution of stars in the sky could produce groups just by
random fluctuations.  In order to test if our detected groups are
special features in the halo or just random fluctuations in a smooth
distribution, we performed extensive simulations of random samples of
stars. The simulated samples have the following properties:

\begin{itemize}

\item Each sample has a random number of stars taken from a
Poisson distribution with mean $N$. The value of $N$ was obtained
by integrating the number density profile of RRLS in the halo with the
limits in area and depth similar to the observed sample. We obtained that
this region of the halo should contain $N_{ab}=42$ from the halo
density profile given in \citet{vivas06} for type ab stars. By assuming that $N_{ab+c}/N_c = 1.29$ \citep{layden95}, 
we obtain a total number of RRLS (all types) of  $N=54$.  
This number
is significantly lower than the number in our sample 
of RRLS (79), which is expected because the \citet{vivas06} profile was
measured in regions that do not contain obvious halo substructures, 
whereas several authors have found evidence of substructures (see
above) in the region under study.  
These substructures are expected to be imprinted on a smooth
background of unrelated RRLS.  
We have used $N=54$ in our simulations to see if random
fluctuations of the background alone can account for any of the groups
identified in Table~\ref{tab-groups}.

\item A distance was assigned to each star by randomly sampling the
  density profile of RRLS given in \citet{vivas06} between 4 and 22 kpc. A random error of 7\%, similar to
the error of our observations, was added.

\item Random coordinates (RA, Dec) were assigned to each star within the limits of the survey area.

\item Radial velocities were drawn from a 
Gaussian distribution with mean 0 ${\rm km~s}^{-1}$ and $\sigma_{los} =$ 116 ${\rm km~s}^{-1}$,
which is the value derived by \citet{brown10} for stars at $\sim 15$ kpc from the Sun.
A random error of 16 ${\rm km~s}^{-1}$ was added to each simulated star.

\end{itemize}

This way, we created 10,000 simulated samples and we detected pairs and groups in each one
of them using exactly the same procedure as with the RRLS sample. Pairs and groups
appeared frequently in the simulated samples.  
There are enough mock groups in our simulations to test statistically
their properties and compare them to our observed groups. 
In the case of $\epsilon=0.046$, for example, there are $\sim36,000$ groups formed
in the simulations (an average of $\sim 3.6$ groups per simulation in 10,000 simulations).
Our goal is to see
if any of the RRLS groups have properties that are not easily reproduced by the random groups.

\begin{figure*}
   \centering
   \includegraphics[angle=-90,width=16cm]{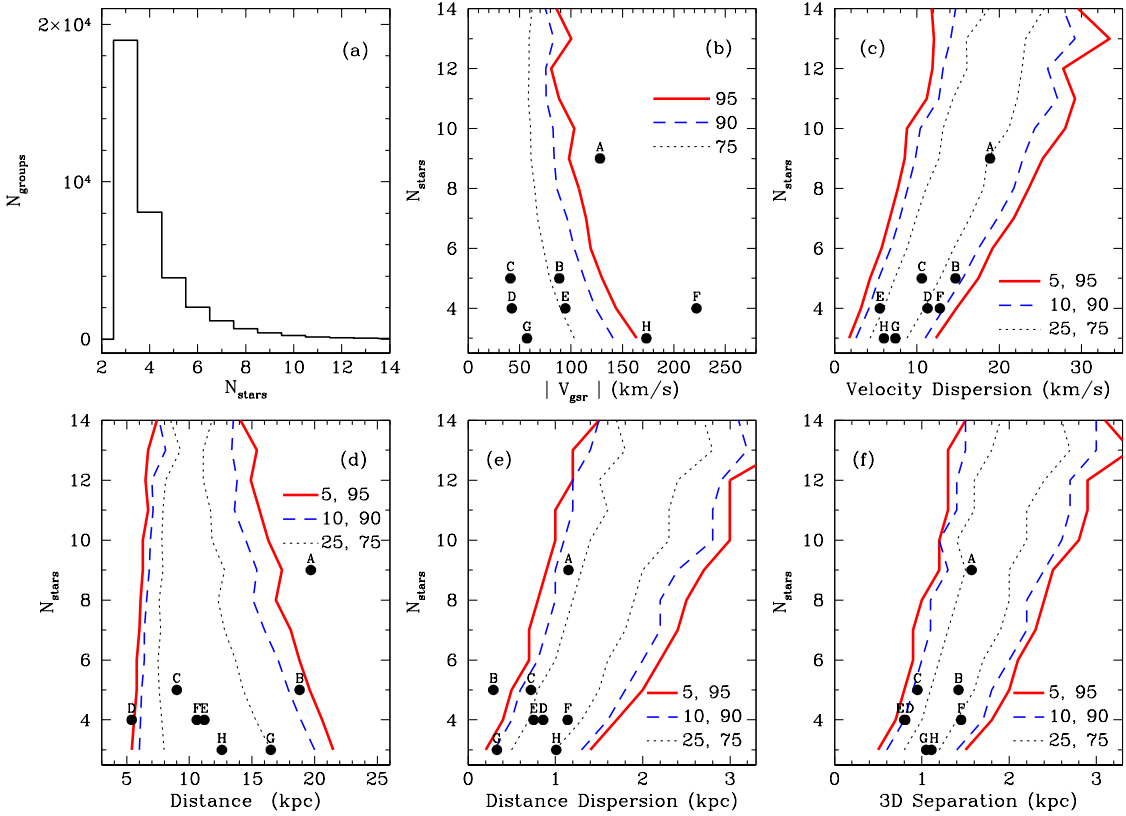}
      \caption{Properties of the groups found in the simulated samples using $\epsilon=0.046$.
      (a) Distribution of the number of stars in the groups. Lines in panels b-f indicate the percentiles
      5, 10, 25, 75, 90 and 95 of the distribution of radial velocity (b), velocity dispersion (c), 
      distance (d), distance dispersion (e) and 3D separation (f), for groups having $N_{stars}$
      members. For the velocity distribution (b) the x axis shows  the absolute value of the
      velocity and hence, only the percentiles 75, 90 and 95 are
      indicated. Panels (b) to (f) show the location of groups A to H
      (Table~\ref{tab-groups}) as solid circles.}
         \label{fig-percentiles}
\end{figure*}

Figure~\ref{fig-percentiles} shows some of the properties of the groups
found in the simulated samples with $\epsilon=0.046$.
Panel (a) shows that most of the mock groups contain only 3
members.  The number of mock groups decreases by a factor of $\sim 2$ for $N_{stars} \geq 4$.
In the groups of RRLS we have 2 groups having 3 members each, 3 groups with 4 members, 
2 with 5 and 1 with 9 members. It is very rare ($<$ 3\%) in our simulations to form a group with 9 or 
more members. Group A with 9 members is hence, significant.

Figure~\ref{fig-percentiles}b shows the absolute value of the mean radial velocity
of the simulated groups. In this plot the lines indicate the
percentiles 75, 90 and 95 in the distribution of velocities of groups with a given N.
Most of the groups in the simulated samples have small velocities 
which agrees with the distribution of velocities given to the random stars. 
Groups with small N may form with relatively large velocities but with
increasing N the largest velocity possible in the groups decreases. The solid points
in Figure~\ref{fig-percentiles}b indicate the velocities of the eight groups of RRLS
indicated in Table~\ref{tab-groups}. Three of those groups (A, F and H)
have velocities which are above the percentile 95 of the distribution. 
Groups F and H have only 4 and 3 members respectively. Although it is very common to have
modelled groups with this low number of stars, it is very rare that they have such a high radial
velocity. This fact makes these groups strong candidates to be real features of the halo and not
just random fluctuations.

The rest of the panels in Figure~\ref{fig-percentiles} shows the distribution of other properties
measured in the random groups, namely, velocity dispersion (c), distance (d), distance dispersion
(e) and mean separation (3D) from the center of the group (f). For these properties we also show
the lines of the percentiles 5, 10 and 25. The mean distance of the groups (Figure~\ref{fig-percentiles}d) seems to be another
parameter that allow us to recognize special features over random fluctuations. At every distance is relatively easy to form groups of small N, but
with increasing distance, the decline in the density of RRLS makes it
increasingly difficult to form groups of large N. Again, group
A, at 19.7 kpc from the Sun, 
seems special since it is above the 95 percentile of the distance distribution for groups with
9 members. In addition, group B (5 members, 18.8 kpc) is above the 90 percentile for groups
of such N.

Group B also stands out because the distance dispersion (Figure~\ref{fig-percentiles}e) is 
significantly smaller than any of the random groups. This group has not a particularly small 3D 
separation, suggesting that the group is more concentrated along the line of sight than across
the sky. This may be understood if the stream is crossing almost perpendicularly our line of sight.

\subsection{Significant Groups}

The analysis of the properties of the simulated groups indicate that some of 
our groups of RRLS cannot be explained as random fluctuations in the distribution of halo
stars. We considered that a group is potentially significant if at least one of their 
properties is below or above the percentiles 10 and 90 of the distribution of simulated
groups for a given N.
This way, five of the groups of RRLS have properties that are quite different from groups formed by 
chance among halo stars. This five groups are thus strong candidates to be real substructures of 
the halo.

Without a doubt, group A is the most significant feature found in this part of the sky. 
First it has a very high number of members, 
which is unusual in the random groups formed in the simulated samples. Furthermore, from the 
few groups formed with 9 members, none has the combination of mean velocity and 
distance  
as group A.  This group corresponds to the VSS as first determined
by \citet{duffau06} and \citet{newberg02}.
The center of the group is found at RA $=186.2\degr$, but it
contains stars basically throughout the region explored, from
$180\degr$ to $195\degr$.

Group B has 5 members and it is also located at a large distance from the Sun, 18.8 kpc.
This is above the 90th percentile of the distance distribution for groups with N=5. Group B
has another remarkable property and it is that it has a very small dispersion along the line
of sight, of only 0.29 kpc, which may be an indication of a thin stream crossing the region
perpendicular to the line of sight.

Groups F and H, with 4 and 3 members respectively, have velocities unusually high 
which make them significant even when having such low number of members.
Group F is the same reported already by \citet{vivas08}.

Finally, group D is unusual because it is formed by the stars closest to the Sun in our sample.
It is remarkable that out of the 5 stars with distances closer than 6.5 kpc, 4 of them make
a tight group. This group is present even with a low value of $\epsilon$ of only 0.035. 

The remaining groups, C, E and G may not be significant and thus are
not discussed further. 

Figure~\ref{fig-percentiles} shows that it would be hard to use
  this method for detecting groups having very low $\vert V_{\rm gsr}
  \vert $ since
they are easily formed among field stars. In order to be detected by
this method, such groups should have an unusual property such as a
very small dispersion in velocity and/or distance.

\subsection{Comparison with other Catalogs and Tracers \label{sec-BHB}}

\begin{figure}
   \centering
   \includegraphics[width=8cm]{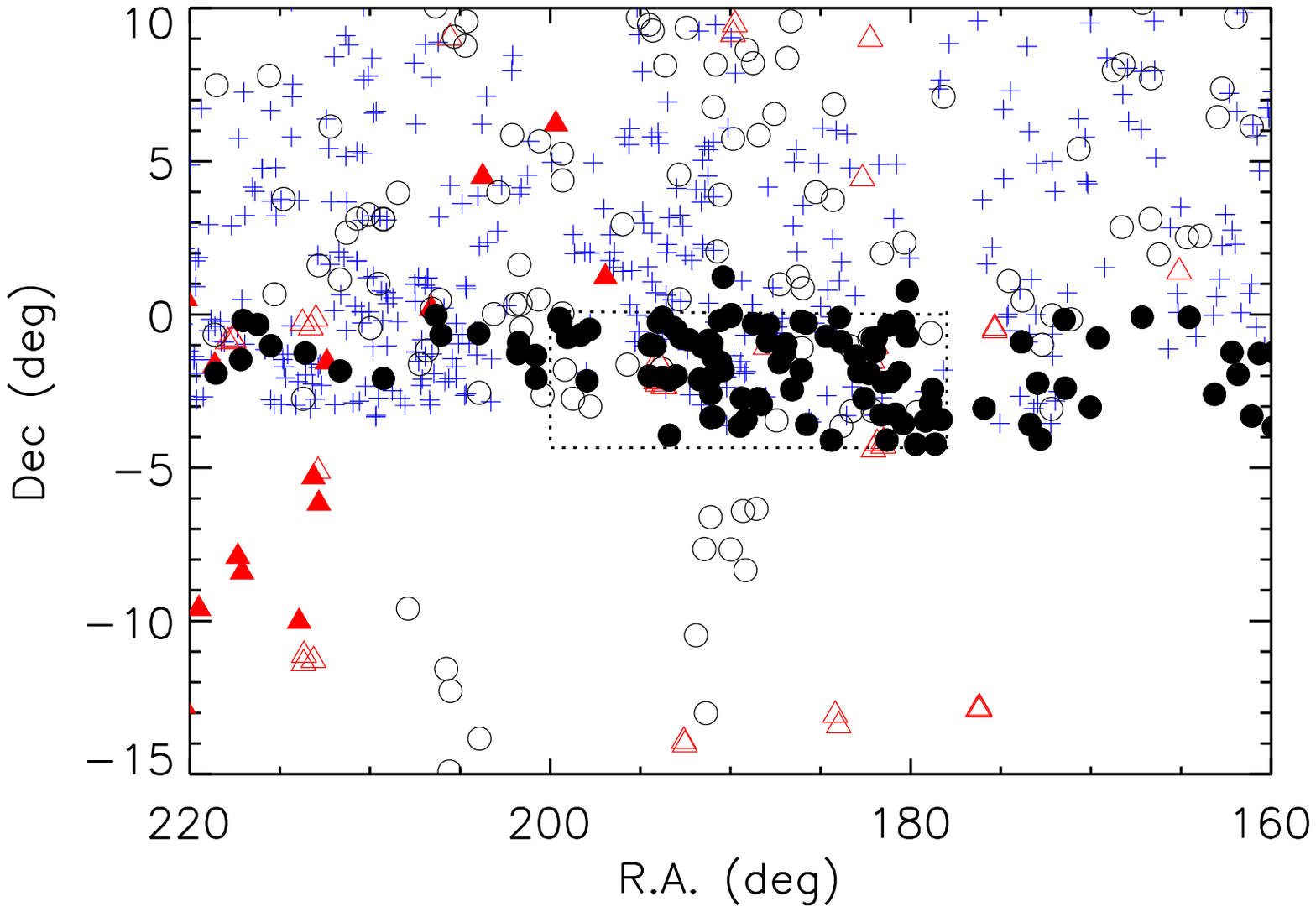}
      \caption{Distribution in the sky near the Virgo region of halo
        tracers for which both velocities and distances are available
        in the literature. Circles represent RRLS (solid circles are
        stars in the QUEST survey, which includes the ones reported in
        this work; open circles come from the SEKBO and
        Catalina surveys), $+$
        symbols are BHB stars from \citet{sirko04a}, \citet{brown08} and \citet{brown10}, open and
        solid triangles are red giants from \citet{starkenburg09} and
        M giants from \citet{majewski04},
        respectively. Notice that the figure shows a region in the sky
        much larger than the one explored in this work with RRLS,
        i.e., the dotted rectangle with $178\degr<$ RA $<200\degr$, $-4\degr<$ Dec $<0\degr$.}
         \label{fig-catalogs}
\end{figure}

Several samples of halo stars with both reliable distances and radial
velocities are available in the literature.  We used those sources to
find additional evidence for the kinematical groups described above.
We used the new spectroscopic catalog of bright
QUEST RRLS, which contains velocity measurements for all the
stars in that catalog brighter than $V\sim16$ (Vivas et al, in
preparation). We included in this catalog the velocity measurement of
QUEST star \#167
made by \citet{casetti09}, which was associated with the VSS by those authors.
We also compared our results with a subset of RRLS in Virgo
from the SEKBO and the Catalina catalogs which have spectroscopic information
\citep{prior09a,drake13}. In the case of Catalina, we found 14 stars in common
with the QUEST catalog and a comparison of the velocities indicates a systematic 
offset of 27 ${\rm km~s}^{-1}$ between both catalogs, with the QUEST velocities being larger than
Catalina's. For the purposes of this work, we added 27 ${\rm km~s}^{-1}$ to all the velocities 
in the Catalina catalog.
We also made use in this study of the latest version of the
catalog of halo red giants in the Spaghetti survey \citep[and
  references herein]{starkenburg09}, as well as M giants from
\citet{majewski04}, most of which are probably associated with the Sgr
dSph galaxy. There are also several BHB catalogs available, namely,
\citet{sirko04a, brown08} and \citet{brown10}.  All of these catalogs
contain stars within or around the region studied in this work, as
shown in Figure~\ref{fig-catalogs}.  We explored also the
catalog of BHB stars by \citet{depropris10} which overlaps our
region. However, these BHB stars are located at distances
$>28$ kpc, and hence they are much farther away than our sample.

In the case of the BHB catalog of \citet{sirko04a} which is based in
SDSS-DR1, we searched for updated radial velocities in SDSS-DR7 in order
to overcome any possible systematic error in the radial velocity in
this early catalog.  In many cases, more than one DR7 spectroscopic
measurement was available, and we calculated an error-weighted mean 
velocity for those stars.  On the other hand, the catalog of
\citet{brown08} is based in 2MASS and, since it is a nearby sample, it
may contain thick disk stars as well as halo stars. Following those
authors, we eliminated all stars with $z<5$ kpc to have a cleaner
sample of halo BHBs. We also made sure all radial velocities were in
the same frame (the galactic standard of rest) as our sample of RRLS.
From Figure~\ref{fig-catalogs}, there are no BHB
data at declinations south of $\sim -3\degr$, thus the SEKBO, Catalina and
Spaghetti samples are specially valuable on those regions.

\begin{figure*}
   \centering
   \includegraphics[width=16cm]{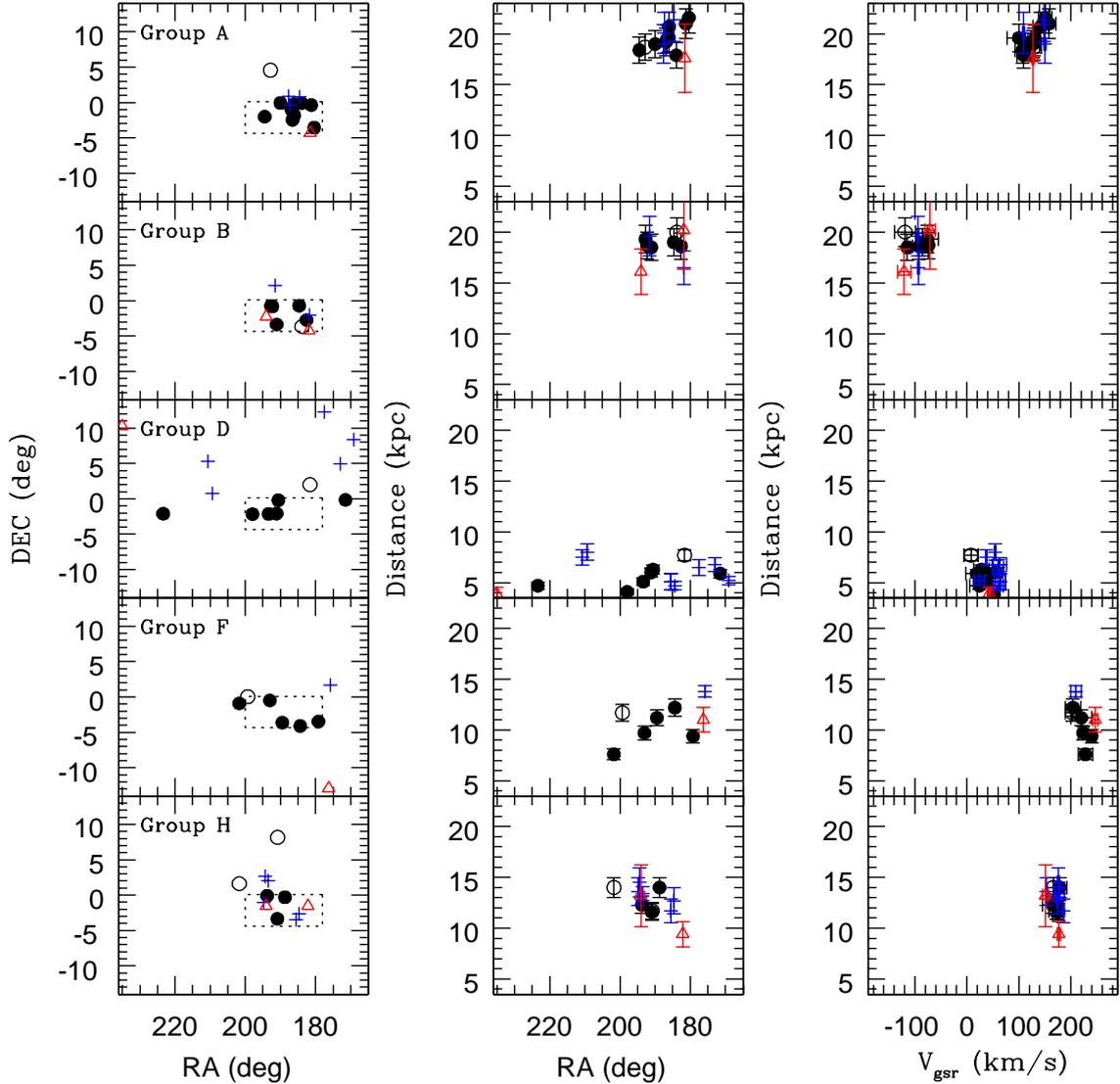}
      \caption{Distribution in the sky and in phase space of the
        groups detected in this work.  Each row shows
        different planes for the same group. Symbols are the same as
        in Figure~\ref{fig-catalogs}.}
         \label{fig-groups_extra}
\end{figure*}

In order to study which other stars are potentially associated with the groups of
RRLS, we determined which RRLS in
our significant groups paired with any other star in the additional
catalogs, using the same algorithm described above with $\epsilon=0.05$.
In this case, however, we did not weight by the errors since the properties of
each catalog are very different from each other.
The results are shown in Figure~\ref{fig-groups_extra}.
Each row in the figure corresponds to different planes (Dec vs
RA, distance vs RA, and distance vs $V_{\rm gsr}$) of the 5
significant groups described in the last section. 
All these groups have RRLS that
pair with several BHB stars and red giants (crosses and triangles
respectively).  Each one of the groups of RRLS have at least one red
giant which is associated with it. Similarly, additional RRLS for each group are provided
by either the SEKBO or Catalina catalogs. 
None of the M giant stars paired with any of our groups. 
The individual members of each one of these groups
are listed in Table~\ref{tab-members}. The table indicates for each
group, the survey from which the member was identified, the type of
star used as tracer in that survey (RRLS: RR Lyrae stars, BHB: Blue
Horizontal Branch stars, and RG: Red Giant stars), the ID within the
survey for the star, its RA and Dec as well as its velocity and
heliocentric distance. Finally, the last column gives the
corresponding reference. Given the very different
  properties of the catalogs used, each one having different coverages
and completeness, we did not attempt to make simulations to test the
significance of those additional members. The additional members suggest possible directions, within the limits of their own surveys, in which follow up observations can focus in the future.

\begin{table*}
\centering
\caption{Individual members of kinematical groups}
\label{tab-members}
\begin{tabular}{lllrccrrl} 
\hline
\hline
Group & Survey & Tracer & ID & RA & Dec & $V_{\rm gsr}$ & d & Reference \\
\hline
           &           &                     &     & (deg) &  (deg) & (${\rm km~s}^{-1}$) & (kpc) &  \\
\hline
A & QUEST & RRLS &                    725 &   180.404037 &    -3.557914 &  151 & 21 & This work \\
A & QUEST & RRLS &                    177 &   181.212601 &    -0.351735 &  158 & 21 & This work \\
A & QUEST & RRLS &                    189 &   183.965347 &    -0.098466 &  109 & 17 & This work \\
A & QUEST & RRLS &                    195 &   186.065384 &    -1.820641 &  142 & 20 & This work \\
A & QUEST & RRLS &                    196 &   186.083054 &    -0.227318 &  100 & 19 & This work \\
A & QUEST & RRLS &                    740 &   186.586151 &    -2.446141 &  131 & 19 & This work \\
A & QUEST & RRLS &                    199 &   186.930542 &    -0.965226 &  129 & 19 & This work \\
A & QUEST & RRLS &                    210 &   190.014893 &    -0.069222 &  127 & 19 & This work \\
A & QUEST & RRLS &                    237 &   194.513031 &    -2.009306 &  107 & 18 & This work \\
A & CSS   & RRLS &   J125124.4+043347 &   192.851690 &     4.563240 &  129 & 18 & \citet{drake13} \\
A & SDSS  & BHB  &                             &   187.650010 &     0.874320 &  150 & 19 & \citet{sirko04a} \\
A & SDSS  & BHB  &                             &   187.396030 &    -0.409560 &  109 & 20 & \citet{sirko04a} \\
A & SDSS  & BHB  &                             &   184.603940 &     0.794670 &  148 & 21 & \citet{sirko04a} \\
A & Spaghetti & RG &               &   181.514723 &    -4.291935 &  127 & 17 & \citet{starkenburg09} \\
\hline 
\end{tabular}
\tablefoot{Only members of group A are shown here. The entire table containing members of all
groups is available in the online version of this article. See text
for description.}
\end{table*}

The left columns in Figure~\ref{fig-groups_extra} show the spatial
distribution of each group (Dec vs RA).  All the groups extend
in the sky to some degree in both directions. It is especially noticeable for group D
and this is just a consequence of that group being close to the Sun.
These plots show us
that the five kinematical groups occupy the same region of the
sky. The right panels in Figure~\ref{fig-groups_extra} show the distribution in phase space for 
every group.  It is clear that each group occupies a very different region
in this plane making the case that they are distinct structures.

 \subsection{Properties of the groups}
 In Table~\ref{tab-properties} we summarized the main properties of
 the kinematical groups we have found.  The mean
 properties were calculated using only the RRLS which are members of
 each group, including the ones from SEKBO, Catalina and from the
 extended QUEST catalog. 
 
 We consider here whether the VSS (group A) may be related with groups F and H. 
 Figure~\ref{fig-groups_extra} shows there is a sequence in the location of these 3 groups
 in phase space (rightmost panels in the Figure), with velocity increasing and distance
 decreasing from group A, to group H, to group F. Table~\ref{tab-properties} indicates that
 the properties of the RRLS in these groups are similar and thus, at this point, a relationship
 among them cannot be discarded. The three groups have similar metallicities, although group A
 has the largest metallicity dispersion among the three. They also have RRab with similar
 mean periods, and  the type ab dominates over the type c among the RRLS in the three groups.
 
 \citet{carlin12} identified turnoff stars with proper motions, in a small region at
 (RA, Dec) = ($178.8\degr$,$-0.6\degr$), in order to calculate an orbit for the VSS. 
 It is worth noticing that the properties of those turnoff stars 
 (mean distance=$14$ kpc, 
 $V_{\rm gsr} =$ $153$ ${\rm km~s}^{-1}$) are more compatible with stars in group H rather than the 
 VSS itself. 
 Ideally, obtaining proper motions in more extended and deeper areas
will allow to determine whether the two groups 
have similar orbits. Curiously, QUEST star \#167, which has been
associated with the 
turnoff stars measured by \citet{carlin12} does not match either
members 
of group H or any other star in the rest of the groups. After
increasing the value of $\epsilon$ to 0.06, 
this star matches 3 members of the VSS (group A) instead.

On the other hand, \citet{newberg07} studied a sample of F stars with bluer colors
than the ones containing the VSS signature.  Among these, one velocity
peak of significant relevance is present on the N07 N field at $\gtrsim
150$ ${\rm km~s}^{-1}$ and at relatively close distances ($<14.5$
kpc).  \citet{newberg07} believed it might be related to the VSS but
did not elaborate more on this.  We can now connect this feature to
our Group H. The presence of this peak on the sample with
colors bluer than the VSS turnoff suggest that it is possible that the
Group H is actually composed of a different population and as such, it
may be a different feature than the VSS.  We notice that this group is not
present on the N07 S field, suggesting that it is not as extended as
the VSS.
 
Group B is peculiar among the groups because is dominated by type c stars (only 1 out of the
6 RRLS in the group belongs to the type ab class). On the other hand, group D has the largest
ratio of BHB/RRLS, although this may be an observational bias given the availability of a large 
catalog of nearby BHB stars \citep{brown08}. Both groups B and D have the shortest mean periods
of the RRab stars among all the groups.
 
 \begin{table*}
 \centering
 \caption{Properties of the groups}
 \label{tab-properties}
 \begin{tabular}{lccccccccccc}
 \hline
 \hline
 Group     & RA        & $N_{RR}$ & $N_{BHB}$ & $N_{BHB}/N_{RR}$ &
 $RR_c/RR_{ab}$ & $P_{ab}$ & [Fe/H] & $\sigma$[Fe/H] & $\langle V_{\rm
   gsr} \rangle$ & $\sigma V_{\rm gsr}$ & $\langle d \rangle$ \\
 \hline
 & (deg) & & & & & (d) & & & (${\rm km~s}^{-1}$) & (${\rm km~s}^{-1}$) & kpc \\
 \hline
 A & 186.863 & 10 & 3   & 0.3 & 0.11 & 0.56 & -1.78 & 0.37 & 128 & 18 & 19.6 \\
 B & 187.879 &   6 & 2   & 0.3 & 5.00 & 0.52 & -1.69 & 0.18 & -94 & 18 & 19.0 \\
 D & 192.792 &  7  & 8   & 1.1 & 0.40 & 0.55 & -1.64 & 0.09 &  32 & 16 &   5.7 \\
 F & 191.194 &   6 & 2   & 0.3 & 0.00 & 0.59 & -1.62 & 0.19 & 220 & 13 & 10.3 \\
 H & 193.187 &   5 & 1   & 0.2 & 0.00 & 0.57 & -1.72 & 0.21 & 172 &  6 & 12.7 \\
 \hline
 \end{tabular}
 \end{table*}

\subsection{Comparison with Sgr models}
 
\begin{figure}
 \centering
 \includegraphics[width=8cm]{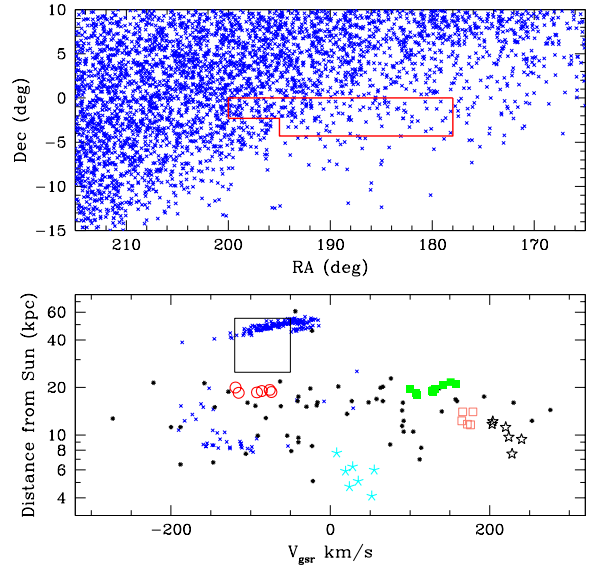}
 \caption{Top: Distribution in the sky around Virgo of the Sgr debris in the models of \citet{law10}. The
 polygon indicates the region studied in this work. Bottom: Velocity and distance of 
 model particles (crosses) and RRLS (asterisks) within the box
 indicated above. Stars in each one of the 5 kinematical groups
 detected here are shown with their corresponding symbol
 (Table~\ref{tab-groups}). The two RRLS at distances $>40$ kpc
 were taken from \citet{vivas05}. Rectangle indicates the approximate location of another kinematical group found by \citet{brink10} and \citet{casey12}.
 }
 \label{fig-sgr}
 \end{figure} 
 
 Figure~\ref{fig-sgr} (top) shows that the models of Sgr debris by \citet{law10} indicate
 that its tidal streams pass close enough to our area of study that we can expect that some
 of our RRLS belong to this galaxy. Although most of the Sgr stars in the region are located
 at distances larger than $\sim$40 kpc (bottom panel in Figure~\ref{fig-sgr}), there are some
 particles at closer distances, within the range studied in this work. However, the nearby debris
 does not seem to be related with any of the kinematical groups found in this work, 
 shown as coloured symbols in Figure~\ref{fig-sgr}. None of our groups have either the velocity 
 or the distance for the expected Sgr debris in the region.

There is a small number of nearby Sgr debris stars in this part of the
sky (distance $\lesssim$ 11 kpc, 
-160 $\lesssim$ $V_{\rm gsr}$ $({\rm km~s}^{-1})$ $\lesssim$ -90). According to \citet{law10} this part of
the stream is made of stars that were mostly stripped from Sgr between
3.2 to 5.0 Gyrs ago. The small number of model particles translates in
an even smaller number of RRLS predicted to exist in that nearby
stream. \citet{zinn14} estimated that for every particle in the model
of \citet{law10} there are $\sim 0.23$ RRLS. Thus, with 24 particles in that nearby stream, there should be $\sim 5$ Sgr RRLS in the region.
While there are 3 RRLS which have similar distance and velocity than the stream (see Figure~\ref{fig-sgr}), 
we also expect, from the simulations described in Section~\ref{sec-sim}, $4\pm2$ halo RRLS in the same region of the phase space. Thus, there is not a clear excess of RRLS that can be readily associated with this stream from an old wrap. A similar conclusion on the lack of strong evidence for a stream stripped between 3.2 to 5 Gyrs ago was made by \citet{zinn14} based on the spatial distribution of a larger sample of RR Lyrae stars.

The bottom panel in Figure~\ref{fig-sgr} also shows the regions where
other kinematical signatures have been found by other
groups. \citet{brink10} and \citet{casey12} studied the velocity
distribution of main sequence stars and red giants in the region,
respectively (see Figure~\ref{fig-spec}). Although \citet{brink10} and
\citet{casey12} state somewhat different ranges for their velocity features, they are consistent with a mean velocity of $\sim -80$ ${\rm km~s}^{-1}$ . This velocity
    is sufficiently close to mean velocity of Group B that they may be
    related.  However, 
the majority of the stars in both the \citet{brink10} and the
\citet{casey12} samples are probably more distant that 20 kpc, and
they may instead be related to the Sgr stream, which lies at $\sim 50$ kpc, or with the substructure recently identified by \citet{jerjen13}) at $\sim 23$ kpc, whose velocity has not yet been measured.

\section{CONCLUSIONS \label{sec-conclusions}}

We present the first complete spectroscopic study of all RRLS in the
QUEST survey in the direction of the VSS. For this work we have
gathered 82 targets combining two previous works by the collaboration
\citep{duffau06,vivas08}, and reporting on 36 RRLS newly observed
stars. The sample covers heliocentric distances between 4-23 kpc,
$\sim 90$ square degrees, and RA between $178\degr$-$200\degr$.
This homogeneous sample enables us to disentangle in phase space several
substructures lying along the same line of sight. This is possible
because RRLS are excellent standard candles and allow the
determination of precise distances. 

 A summary of all kinematic groups found in the Virgo region is displayed in
 Table~\ref{comp-tab}. In this table, each check mark ($\surd$) under
 a study reference represents a detection of the corresponding group.
 The corresponding references to each study are: N07
 \citep{newberg07}, B10 \citep{brink10}, C12 \citep{casey12}, S09
 \citep{starkenburg09} and P09 \citep{prior09a}.  The first column
 contains, in addition to the 5 groups of RRLS detected here, two
 groups predicted to lie in this region by the models of
 the disruption of Sgr (``Sgr far'' refers to stars located at distances
 $>40$ kpc, while ``Sgr near'' correspond to stars between $\sim 8$ and 18
 kpc, as shown by the debris (asterisks) in Figure~\ref{fig-sgr}).  An
 asterisk in Table~\ref{comp-tab} indicates a weak detection of the corresponding group.

\begin{table*}
 \centering
\caption{Kinematic Detections in the Virgo Region}
\label{comp-tab}
\begin{tabular}{lcccccc}
\hline
\hline
Group & This work & N07 & B10 & C12 & S09 & P09 \\
 & RRLS & F-TO  & F-TO & K-giants & K-giants & RRLS \\
\hline
Group A (VSS) &  $\surd$  & $\surd$ & $\surd$  & $\surd$ &                &  $\surd$ \\ 
Group B          &  $\surd$  &     *        & $\surd$  &               &                &               \\
Group D         &  $\surd$  &               &               &               &                &                \\
Group F          &  $\surd$  &               & $\surd$ &  $\surd$ &                &                \\
Group H          &  $\surd$ & $\surd$ &                &               & $\surd$   &               \\
Sgr far             &                &               & $\surd$ & $\surd$ &                 &               \\
Sgr near          &                & $\surd$  & *            &                & $\surd$  & $\surd$ \\
\hline
\end{tabular}
\end{table*}

The VSS, which is now reconciled with feature S297+63-20.5 \citep{newberg07},
is the most significant feature in our sample. It spans at
least between 17.9-21.6 kpc along the line of sight and has a mean radial
velocity of 128 ${\rm km~s}^{-1}$, as deduced from its RRLS components. 
A possible relationship of the VSS with two other groups (Groups F and H) cannot be 
discarded at this point. The three groups form a sequence in
  phase space with the VSS and group F at the two ends of the sequence
with group H intermediate in both velocity and distance (see
Figure~\ref{fig-groups_extra} and Figure~\ref{fig-sgr}).

Another significant group, composed mostly of type c stars, was found at a similar
distance than the VSS, 19 kpc, but with a velocity of $-94$ ${\rm km~s}^{-1}$.
Similar velocity signatures have been found in the region by \citep{brink10} and
\citet{newberg07}, although the latter found it among the brightest stars in their sample, hence,
presumably corresponding to a group at a shorter distance than our detection.
Finally, a nearby group at only 6 kpc was also detected with a mean velocity of 
$V_{\rm gsr}=32$ ${\rm km~s}^{-1}$.
The properties of this group are harder to isolate due to its
much larger angular extension in the sky and its velocity being close to the expected mean velocity
of halo stars.

Using the models of \citet{law10}, we find that while the Sgr stream debris are not expected to be 
dense in this region at distances closer than 40 kpc, some debris is expected to be found 
in the VSS area. 
Several of our RRLS lie within the expected distance and velocity of those predictions. 
However, none of the kinematic groups we found in this work seem to be related with Sgr debris.
For the distances involved in this study the dominant feature is the VSS. 
This was not the case in the works of \citet{brink10} and \citet{casey12}. Their exploration included 
targets presumably at much larger distances. In those cases, the dominant feature seems to be
the Sgr leading stream or another new sub-structure in the region as suggested by \citet{jerjen13}.

The numerous kinematic features found in our sample of RRLS
qualitatively agree with
the expectations of halo substructures in cosmological models of
galaxy formation.  Our results suggest that several streams,
rather than a single accretion event, are responsible for the excess
of stars found in the Virgo region.

The true shape and extension of all the streams detected in this work
will require accurate distances and velocities of targets in a more
extended region than the one considered here.  The possibility of
clarifying which of the different substructures are related
to each other and which are not, might require additional information.
One possibility is the chemical tagging of their
stars, which is one direction we are currently pursuing.

\begin{acknowledgements}
 We thank the anonymous referee for useful suggestions and comments. We thank Amina Helmi for kindly providing us the data of the red giants found
 in the Spaghetti survey. We thank Dana Casetti-Dinescu for kindly providing her
 orbit for RRL 167.
 S.D. acknowledges the support from both ARI and CIDA during her visit to
 CIDA, Venezuela in November 2010 to work in this project, and to Yale/U. de
 Chile through the joint PhD program (MECESUP UCH0118,
Fundacion Andes C13798, Fondap 15010003 and CONICYT 24070078. S.D.
also acknowledges support from Sonderforschungsbereich
SFB 881 “The Milky Way System” (subprojects A2, A3 and A4) of the German Research Foundation (DFG).
R.Z. acknowledges support from National Science Foundation grants AST 05-07364
 and AST 11-08948. M.T.R and R.M. received partial support from CATA (PB06)
 CONICYT. A.K.V. thanks the hospitality of the Department of Astronomy at University of 
 Michigan during her sabbatical leave in which part of this work was made.

This research was based on data obtained at KPNO (Kitt Peak National
Observatory, National Optical Astronomy Observatory, which is operated by the
Association of Universities for Research in Astronomy (AURA) under cooperative
agreement with the National Science Foundation), WIYN (The WIYN Observatory is
a joint facility of the University of WIsconsin-Madison, Indiana University,
Yale University, and the National Optical Astronomy Observatory), CTIO (Cerro
Tololo Inter-American Observatory, National Optical Astronomy Observatory,
which is operated by the Association of Universities for Reseach in Astronomy,
under contract with the National Science Foundation), telescopes at the
European Southern Observatory La Silla, Magellan's Clay telescope at LCO (operated by a
consortium consisting of the Carnegie Institution of Washington, Harvard
University, MIT, the University of Michigan, and the University of Arizona) and instruments operated
by the SMARTS Consortium. We thank the staff at all of these observatories for
their help during the time the observations were conducted.
\end{acknowledgements}

\bibliographystyle{aa} 
\bibliography{mybib.bib}

\end{document}